\documentclass{ametsoc}

\journal{mwr}

\bibpunct{(}{)}{;}{a}{}{,}

\title{A new dynamical core of the {G}lobal {E}nvironmental {M}ultiscale ({GEM}) model with a height-based terrain-following vertical coordinate}

\authors{Syed Zahid Husain$^1$\correspondingauthor{Syed Zahid Husain, Atmospheric Numerical Prediction Research Section, Meteorological Research Division, Environment and Climate Change Canada, Dorval, QC H9P 1J3.}, Claude Girard$^1$, Abdessamad Qaddouri$^1$ and Andr\'{e} Plante$^2$}

\affiliation{{$^1$Atmospheric Numerical Prediction Research Section, Meteorological Research Division, Environment and Climate Change Canada, Dorval, QC H9P 1J3\\}
{$^2$Canadian Meteorological Centre,\\Environment and Climate Change Canada, Dorval, QC H9P 1J3}}

\email{Syed.Husain@canada.ca}

\abstract{A new dynamical core of Environment and Climate Change Canada's Global Environmental Multiscale (GEM) atmospheric model is presented. Unlike the existing log-hydrostatic-pressure-type terrain-following vertical coordinate, the proposed core adopts a height-based approach. The move to a height-based vertical coordinate is motivated by its potential for improving model stability over steep terrain, which is expected to become more prevalent with the increasing demand for very high resolution forecasting systems. A dynamical core with height-based vertical coordinate generally requires an iterative solution approach. In addition to a three-dimensional iterative solver, a simplified approach has been devised allowing the use of a direct solver for the new dynamical core that separates a three-dimensional elliptic boundary value problem into a set of two-dimensional independent Helmholtz problems. The new dynamical core is evaluated using numerical experiments that include two-dimensional nonhydrostatic theoretical cases as well as 25-km resolution global forecasts. For a wide range of horizontal grid resolutions---from a few meters to up to 25 km---the results from the direct solution approach is found to be equivalent to the iterative approach for the new dynamical core. Furthermore, results from the numerical experiments confirm that the new height-based dynamical core leads to results that are equivalent to the existing pressure-based core.}

\usepackage[italicdiff]{physics}
\usepackage{enumerate}
\begin{document}

%% Necessary!
\maketitle

%%%%%%%%%%%%%%%%%%%%%%%%%%%%%%%%%%%%%%%%%%%%%%%%%%%%%%%%%%%%%%%%%%%%%
% MAIN BODY OF PAPER
%%%%%%%%%%%%%%%%%%%%%%%%%%%%%%%%%%%%%%%%%%%%%%%%%%%%%%%%%%%%%%%%%%%%%
%

\section{Introduction} \label{s_intro}
The dynamical core of the Global Environmental Multiscale (GEM) model, used operationally by Environment and Climate Change Canada (ECCC) for numerical weather prediction (NWP), employs a log-hydrostatic-pressure-type terrain-following vertical coordinate. The system of nonlinear model equations is linearized around a basic state and is then reduced to an elliptic boundary value (EBV) problem through numerical discretization and elimination of variables \citep{gdm14}. The existing pressure-type coordinate system then permits the use of a direct solver for the discretized EBV problem to resolve the dynamical component of the flow. The direct solver starts by separating the EBV problem vertically in terms of the vertical eigenvectors of the part of the coefficient matrix that only includes the discretized difference and average operators in the vertical direction \citep{qle10}. For $N$ number of model vertical levels, the resulting $N$ vertically-decoupled two-dimensional Helmholtz problems are then separated along the longitude, leading to a system of tridiagonal problems depending only on the latitude. The tridiagonal problems are finally solved using LU decomposition without pivoting. Such an approach is computationally more efficient than most iterative methods, particularly for the spatial resolutions of the current operational NWP systems at ECCC.\\

One of the principal incentives for the adoption of the existing pressure-type vertical coordinate in GEM is the computational advantage of the direct solution approach that is permissible with such a coordinate. However, numerical experiments carried out at ECCC have revealed that vertical separability, which is an imperative for the direct solution approach, can become quite restrictive for very high spatial resolutions, e.g., for sub-kilometer horizontal grid spacing. Furthermore, the reduction of the 2D Helmholtz problems into the 1D tridiagonal problems requires projection of right hand side (RHS) of the 2D problems along the pertinent eigenvectors which is based on Fourier transformation. This necessitates transposing the coefficient matrix that involves global communications, and therefore, becomes inefficient for very large number of processor cores. As a result, the direct solver is found to lose scalability with increasing number of processor cores. Initial research at ECCC as well as published research literature \citep{msc14} reveal that optimized three-dimensional iterative solvers may possess better scalability in these circumstances. The different limitations of the existing direct solver, particularly its potential lack of scalability for future generations of massively parallel supercomputers, therefore warrants the development of more scalable iterative solvers at ECCC. A height-based vertical coordinate is considered to be more amenable to such iterative solvers as the metric terms originating from the vertical coordinate transformation appear explicitly in the discretized EBV.\\

Another, but more challenging, problem pertaining to the existing GEM dynamical core is the fact that the  current model exhibits strong numerical instability over steep orography. Tests conducted at ECCC have determined the maximum permissible terrain slope for maintaining stability to be approximately 45$^\circ$ \citep{vbg15}. This is generally considered to be a limitation inherent to the terrain-following coordinate (TFC) systems \citep{zan12}. With a growing demand for very high-resolution operational NWP systems, steep orographic slopes are expected to become more prevalent in the near future. Improving model stability over steep mountains is therefore of critical importance for the future development of sub-kilometer NWP systems. A number of approaches have been investigated to improve numerical stability with the existing pressure-type vertical coordinate in GEM. These include increased off-centering in the discretized vertical momentum equation, a vertically-variable basic state temperature profile, and modifications to the nonhydrostatic contributions in the linear system arising from the discretized GEM formulation. However, none of these approaches has been found to lead to any meaningful stability improvement for steep orography.\\

Although the instability induced by steep orography is often characterized as a limitation of the terrain-following nature of the vertical coordinate itself, \cite{ssw07} were successful in resolving flow around the Pentagon with a model involving height-based TFC where the maximum slope was well above the widely acknowledged 45$^\circ$ threshold. Previous experience with the Mesoscale Compressible Community model at ECCC also suggests that a dynamical core with height-based TFC does not suffer from similar severe orography-induced instability \citep{gbd05}. Apparently, a dynamical core with a height-based TFC can lead to improved numerical stability through better implicit treatment of the metric terms arising from the vertical coordinate transformation through iterative solvers. More importantly, conventional numerical approximation of the horizontal gradients in a TFC becomes less accurate with increasing terrain slope as well as with increasing vertical resolution close to the model surface \citep{mah84}. In this context, \cite{zan12} argues that the pressure gradient term, in particular, becomes susceptible to triggering numerical instability when the height difference between two adjacent grid points along a terrain-following vertical level is much larger than the vertical grid resolution adjacent to the level. Numerical approximation of the horizontal gradient terms in the TFC, however, can be significantly improved following the corrections proposed by \cite{mah84}. These corrections require determination of the modified horizontal differencing stencils associated with each grid-point location that minimize the error in the metric corrections for the terrain-following nature of the coordinate. The existing pressure-type TFC varies with time and, therefore, would require determination of the pertinent grid-point locations in the vertical for the modified differencing stencils at every time step. On the contrary, the height-based TFC is time-invariant and thus would require the determination of these grid-point locations only once at the beginning of the time integration.  Therefore, from a computational efficiency standpoint, a height-based TFC is more suitable for implementing improved numerical approximation of horizontal gradients to address instabilities induced by steep orography.\\

The aforementioned challenges associated with the existing log-hydrostatic-pressure-type TFC motivated the development of a new dynamical core for the GEM model that utilizes a height-based TFC. The primary objective of the present study is to demonstrate that, for the model configurations where orography-induced numerical instability is not relevant---i.e., for horizontal grid resolutions within the hydrostatic regime---the new dynamical core developed at ECCC with height-based TFC makes predictions that are equivalent to those from the existing model. The present study also explores the appropriate strategy for coupling the operational Physics Parameterization Package (PPP) of RPN (Recherche en pr\'{e}vision num\'{e}rique) with the new height-based dynamical core. Different setups for numerical experiments are utilized to compare the newly-developed dynamical core with the existing one covering both hydrostatic and nonhydrostatic scenarios. The experiments include two-dimensional theoretical cases \citep{rob93,slf02} as well as three-dimensional global forecasts over a Yin-Yang grid \citep{qle11}.\\

Relevant background information on the GEM dynamical core with the proposed height-based TFC---from the spatial and temporal discretizations to the derivation of the elliptic boundary value problem---is presented in section \ref{s_mod_descri}. The different solution methods utilized for the discretized elliptic problem is discussed in section \ref{s_solver}. The issue of coupling between the dynamical core and the parameterized physics forcings is presented in section \ref{s_coupling}. Section \ref{s_evaluation} contains the comparisons between the existing and the proposed dynamical cores in the context of two-dimensional theoretical benchmark cases as well as three-dimensional deterministic global predictions. The conclusions are then summarized in section \ref{s_summary}.\\

\section{Model Description} \label{s_mod_descri}

\subsection{Governing equations}
The GEM model equations originate from the Euler equations. With the traditional shallow atmosphere approximation \citep{phi66}, the system of equations in a spherical coordinate $(\lambda,\phi,r)$ can be expressed as follows:

\begin{equation}
 \dv{u}{t}-\left(f+\frac{\tan\phi}{a}u\right)v+\frac{1}{\rho}\pdv{p}{x} = \left(\dv{u}{t}\right)_{phys},
 \label{e_u_mom}
\end{equation}
\begin{equation}
 \dv{v}{t}+\left(f+\frac{\tan\phi}{a}u\right)u+\frac{1}{\rho}\pdv{p}{y} = \left(\dv{v}{t}\right)_{phys},
 \label{e_v_mom}
\end{equation}
\begin{equation}
 \dv{w}{t}+\frac{1}{\rho}\pdv{p}{z} +g = \left(\dv{w}{t}\right)_{phys},
 \label{e_w_mom}
\end{equation}
\begin{equation}
 \dv{\ln \rho}{t}+\pdv{u}{x}+\pdv{v}{y}+\pdv{w}{z}-\frac{\tan\phi}{a}v=\left(\dv{\ln \rho}{t}\right)_{phys},
 \label{e_conti}
\end{equation}
\begin{equation}
 \dv{\ln T}{t}-\kappa\dv{\ln p}{t}=\left(\dv{\ln T}{t}\right)_{phys},
 \label{e_thermo}
\end{equation}
where Eqs. (\ref{e_u_mom})--(\ref{e_thermo}) govern the evolutions of the $u$, $v$, and $w$ components of velocity,  mass and energy, respectively. The spatial coordinates in the above equations are denoted by $(x,y,z)$ which are related to the spherical coordinate $(\lambda,\phi,r)$ through the differential relations given by
\begin{equation}
 dx=a\cos\phi d\lambda, dy=ad\phi, dz=dr,
 \label{e_spherical}
\end{equation}
such that $u$, $v$ and $w$ are the physical wind components. In Eq. (\ref{e_spherical}), $a$ denotes Earth's radius. The Lagrangian derivative in this case can be expressed as
\begin{equation}
 \dv{}{t}=\pdv{}{t}+u\pdv{}{x}+v\pdv{}{y}+w\pdv{}{z}.
 \label{e_lagrange}
\end{equation}
In addition to the four independent variables $(x,y,z,t)$, the system of five equations (\ref{e_u_mom})--(\ref{e_thermo}) involves six dependent variables, namely, the velocity components $u$, $v$, and $w$, the temperature $T$, the pressure $p$, and the density $\rho$. Also, in the above equations, $f$ is the Coriolis parameter and $\kappa=R/c_p$ where $R$ is the gas constant and $c_p$ is the specific heat of air at constant pressure. The terms on the RHS of Eqs. (\ref{e_u_mom})--(\ref{e_thermo}) with subscript ``$phys$''  denote the various physical forcings. Depending on the equation, these physical forcings may arise from different sources that include friction, diabatic heating and frictional dissipation of kinetic energy. A sixth equation is required to close the system described by the six dependent variables and is provided by the equation of state, given by
\begin{equation}
 p=\rho RT.
 \label{e_state}
\end{equation}

It is important to note that the atmospheric substance does not only contain dry air but also water vapor and different types of hydrometeors. The displacement and evolution of water vapor and hydrometeors in the atmosphere are governed by their own evolution equations. However, they will also affect the RHS terms through fluxes of water vapor and precipitation which constitute sources of mass. The total air density in the presence of water in its different forms can be expressed as
\begin{equation}
 \rho=\rho_d+\rho_w=\rho_d(1+r_w),
 \label{e_rho}
\end{equation}
where $\rho_d$ is the dry air density, $\rho_w=\frac{\rho_w}{\rho_d}$ is the density of water vapor and hydrometeors, and $r_w$ is the mixing ratio for the total water content of the atmosphere. The equation of state in such a scenario is strictly given by
\begin{equation}
 p=(\rho_dR_d+\rho_vR_v)T,
 \label{e_state_2}
\end{equation}
where $\rho_v$ and $R_v$ are the density and gas constant of water vapor. Eq. (\ref{e_state_2}) can then be further rearranged in terms of the total air density $\rho$ as
\begin{equation}
 p=\rho R_d T_v,
 \label{e_state_3}
\end{equation}
where $T_v$ is the virtual temperature of moist air which is given by
\begin{equation}
 T_v=\frac{1+\frac{R_v}{R_d}r_v}{1+r_w}T,
 \label{e_tv}
\end{equation}
where $r_v=\frac{\rho_v}{\rho_d}$ is the water vapor mixing ratio. Rewriting the dynamical equations (\ref{e_u_mom})--(\ref{e_thermo}) in terms of $T_v$ is helpful as the equations can then be expressed using only the dry gas constant that does not vary due to the atmospheric water content. It also allows to account for the effects of water vapor buoyancy and condensed water loading implicitly. Furthermore, from the adiabatic point of view, the introduction of $T_v$ has no consequence since the water content is conserved during dynamical transport. \\

\subsection{Vertical coordinate}
The log-hydrostatic-pressure-type terrain-following vertical coordinate of the operational GEM model \citep{gdm14} has the form
\begin{equation}
\ln\pi=\xi+Bs,
 \label{e_vgd_p}
\end{equation}
where $\xi$ defines the terrain-following vertical coordinate, $\pi$ denotes the hydrostatic pressure, $B$ is a metric term prescribing the rate of flattening of the vertical coordinate with
elevation, and $s=\ln(\pi_s/p_{ref})$ with $\pi_s$ being the hydrostatic pressure at the surface and $p_{ref}=10^3$ hPa is a reference pressure. The definition of this vertical coordinate follows the concept of the generalized hydrostatic-pressure-type hybrid coordinate proposed by \cite{lap92}. Further details regarding the log-hydrostatic-pressure-type vertical coordinate are provided by \citet{gdm14} and \citet{hgi17}.\\

The present study proposes to develop a GEM dynamical core where the vertical coordinate, given by Eq. (\ref{e_vgd_p}), in the existing dynamical core is replaced by a height-based TFC. The traditional formulation for height-based TFC can be expressed as
\begin{equation}
\zeta(z)=H\frac{z-z_S}{z_T-z_S},
 \label{e_vgd_z1}
\end{equation}
where $z$ is the the true geometric height, $z_S$ and $z_T$ are the surface and the model top level heights, and $H$ is a scaling constant. A more general form of Eq. (\ref{e_vgd_z1}) can be devised as
\begin{equation}
z=A+Bz_S,
 \label{e_vgd_z2}
\end{equation}
where $A=(z_T/H)\zeta$ and $B=(1-\zeta/H)$. Assigning $H=z_T$ implies $z_T=\zeta_T$ and, as a result, Eq. (\ref{e_vgd_z2}) becomes
\begin{equation}
z=\zeta+Bz_S
 \label{e_vgd_z3},
\end{equation}
which is similar to Eq. (\ref{e_vgd_p}) in form. The vertical coordinate for the proposed dynamical core in the present study, however, is further generalized as
\begin{equation}
z=\zeta+B_1z_{SL}+B_2(z_S-z_{SL}),
 \label{e_vgd_zs1}
\end{equation}
which follows the concept of SLEVE (Smooth LEvel VErtical)-like coordinate system proposed by \cite{slf02}, where $z_{SL}$ denotes the large-scale components of the orography. The vertical coordinate defined by Eq. (\ref{e_vgd_zs1}) permits separate rates of flattening for the large and small-scale contributions of the orography on the terrain-following vertical coordinate with changing elevation through the metric terms $B_1$ and $B_2$ that are defined as
\begin{equation}
B_n=\bigg(\frac{\zeta_T-\zeta}{\zeta_T-\zeta_S}\bigg)^{r_n},
\label{e_vgd_zbn}
\end{equation}
where $r_n=[r_{n,max}-(r_{n,max}-r_{n,min})\lambda_k]$ and $\lambda_k=(\zeta_1-\zeta_k)/(\zeta_1-\zeta_S)$. The values of $r_{n,min}$ and $r_{n,max}$ together determine the rate of flattening of the vertical levels with increasing height. The subscript $k$ of $\lambda$ indicates the model vertical level number. Furthermore, the value of $k$ decreases with increasing height above the surface such that $k$=1 indicates the top-most model level.\\

Henceforth, in this paper, the two dynamical cores with vertical coordinates based on log-hydrostatic-pressure and height are referred to as GEM-P and GEM-H, respectively. Different aspects of the GEM-P dynamical core, including the model formulation, discretization and numerical solution of the discretized problem along with the various modifications to the formulation over the past years, have been discussed in detail in the existing literature \citep{ycg02,qle11,gdm14,hgi17}. The following subsections therefore only present the relevant details of the proposed GEM-H dynamical core.\\

\subsection{GEM-H formulation}
The development of the GEM-H formulation requires further modifications to the system of equations (\ref{e_u_mom})--(\ref{e_thermo}). First, virtual temperature, given by Eq. (\ref{e_tv}), is introduced in the system of equations along with an isothermal basic state temperature $T_*$ such that $T_v=T'_v+T_*$, where $T'_v$ is the temperature deviation. The corresponding basic state pressure $p_*$ is defined hydrostatically as $\partial(\ln p_*)=-g\partial{z}/(R_dT_*)$. The equation of state, given by Eq. (\ref{e_state_3}), is then used to eliminate density $\rho$ as a prognostic variable followed by a transformation of the resulting equations from the geometric height coordinate to the terrain-following $\zeta$-coordinate. The vertical coordinate transformation leads to the replacements of the independent variables $(x,y,z)$ associated with the $z$-coordinate by $(X,Y,\zeta)$ that are defined in the $\zeta$-coordinate. As a result, the system of equations (\ref{e_u_mom})--(\ref{e_thermo}) is modified as follows:
\begin{equation}
 \dv{u}{t}-\bigg(f+\frac{\tan\phi}{a}u\bigg)v+\frac{T_v}{T_*}\bigg(\pdv{q}{X} -\frac{J_X}{J_{\zeta}}\pdv{q}{\zeta}\bigg)= \left(\dv{u}{t}\right)_{phys},
 \label{e_u_1}
\end{equation}
\begin{equation}
 \dv{v}{t}+\bigg(f+\frac{\tan\phi}{a}u\bigg)u+\frac{T_v}{T_*}\bigg(\pdv{q}{Y} -\frac{J_Y}{J_{\zeta}}\pdv{q}{\zeta}\bigg)= \left(\dv{v}{t}\right)_{phys},
 \label{e_v_1}
\end{equation}
\begin{equation}
 \dv{w}{t}+\frac{T_v}{T_*}\bigg(\frac{1}{J_{\zeta}}\pdv{q}{\zeta}-g\frac{{T_v}^{'}}{T_v}\bigg)= \left(\dv{w}{t}\right)_{phys},
 \label{e_w_1}
\end{equation}
\begin{equation}
 \dv{}{t}\bigg(\frac{q}{c_*^2}+\ln{J_\zeta}\bigg)+\pdv{u}{X}+\frac{1}{\cos\phi}\pdv{(\cos\phi{v})}{Y}+\pdv{\dot{\zeta}}{\zeta}-\frac{g}{c^{2}_{*}} w =\left(\dv{\ln \rho T_v}{t}\right)_{phys}
 \label{e_c_1},
\end{equation}
\begin{equation}
 \dv{}{t}\bigg[\ln\bigg(\frac{T_v}{T_*}\bigg)-\frac{q}{c_{pd}T_*}\bigg]+\frac{N^{2}_{*}}{g}w=\left(\dv{\ln T_v}{t}\right)_{phys},
 \label{e_t_1}
\end{equation}
where $q=R_dT_*\ln(p/p_*)$ is the pressure deviation from the basic-state pressure $p_*$, $\dot{\zeta}=\dv{\zeta}{t}$ is the vertical motion with respect to the transformed $\zeta$-coordinate, $N^2_*=g^2/(c_{pd}T_*)$ is the square of the basic-state Brunt-V\"{a}is\"{a}l\"{a} frequency and $c^2_*=R_dT_*/(1-\kappa_d)$ is the square of the speed of sound. In Eq. (\ref{e_c_1}), $\kappa$ is replaced by $\kappa_d=R_d/c_{pd}$ as an approximation. It is also important to note that, the physical forcings associated with the modified continuity equation, given by Eq. (\ref{e_c_1}), now includes the same diabatic heating term that appears in the thermodynamic equation, given by Eq. (\ref{e_t_1}).

In the above equations, the terms $J_X$, $J_Y$ and $J_\zeta$ appears due to the vertical coordinate transformation where $J_X=\pdv{z}{X}$, $J_Y=\pdv{z}{Y}$ and $J_\zeta=\pdv{z}{\zeta}$. It is however important to note that the coordinate transformation used to derive the Eqs. (\ref{e_u_1})--(\ref{e_t_1}) is incomplete and only the following derivative operators have been transformed:
\begin{equation}
 \pdv{}{x}=\pdv{}{X}-\frac{J_X}{J_\zeta}\pdv{}{\zeta},
 \label{e_trans_x}
\end{equation}
\begin{equation}
 \pdv{}{y}=\pdv{}{Y}-\frac{J_Y}{J_\zeta}\pdv{}{\zeta},
 \label{e_trans_y}
\end{equation}
\begin{equation}
 \pdv{}{z}=\frac{1}{J_\zeta}\pdv{}{\zeta},
 \label{e_trans_z}
\end{equation}
\begin{equation}
 \dv{}{t}=\pdv{}{t}+u\pdv{}{X}+v\pdv{}{Y}+\dot{\zeta}\pdv{}{\zeta}.
 \label{e_lagrange_zeta}
\end{equation}
As a result, the original vertical velocity $w$ has not been completely eliminated from the system. The system of equations in the $\zeta$-coordinate has its own vertical velocity in the form of $\dot{\zeta}=\dv{\zeta}{t}$, whereas $w=\dv{z}{t}$ remains in the system as a kinematic relation that needs to be dealt with explicitly. \cite{czg04} have demonstrated that such an approach is perfectly equivalent to a full coordinate transformation. As the treatment of advection in GEM is based on the semi-Lagrangian approach \citep{hgi17}, the kinematic relation defining $w$ is also solved semi-Lagrangially. However, for convenience, the kinematic relation is modified as 
\begin{equation}
\frac{d}{dt}(z-\zeta)+\dot{\zeta}-w=0,
 \label{e_k_1}
\end{equation}
where Eq. (\ref{e_k_1}) along with the Eqs. (\ref{e_u_1})--(\ref{e_t_1}) constitute the complete system of equations for the GEM-H formulation.

One important aspect of any NWP model is how the effects of the physics forcings, as presented in the RHS of the Eqs. (\ref{e_u_1})--(\ref{e_t_1}), are accounted for as the model equations are integrated at each time step. One way is to resolve the dynamical equations in the absence of any physics forcing and then modify the solution with the parameterized forcing as adjustments outside the dynamics step. Another possible approach is to compute the physics forcing and combine their effects with the nonlinear terms in a semi-implicit way within the dynamics step. This important aspect of the GEM model with particular focus on its impact pertaining to the GEM-H dynamical core is discussed in further details in section \ref{s_coupling}.\\

\subsection{Spatial grid and discretization}
\begin{figure}
\noindent\includegraphics[width=45pc,angle=0]{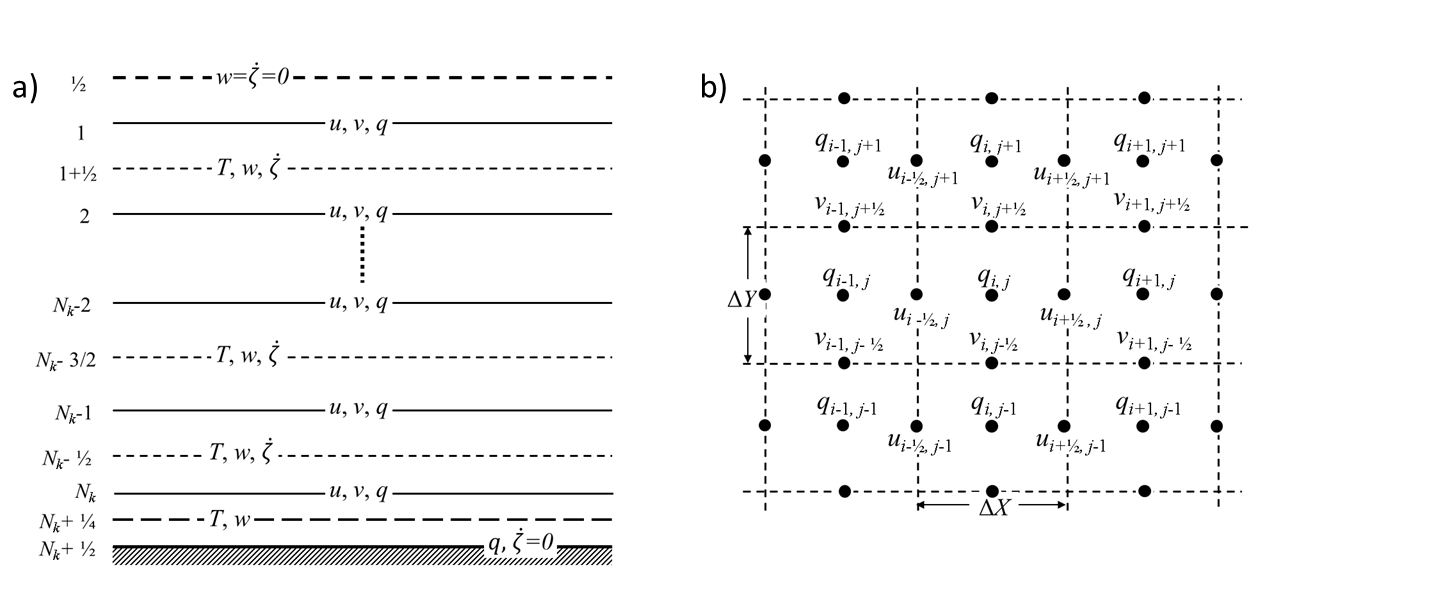}\\
\caption{a) Vertical Charney-Phillips grid. b) Horizontal Arakawa C grid.}
\label{f_grid}
\end{figure}

The objective of this project has been to implement the option of a dynamical core based on height-type vertical coordinate in addition to the existing pressure-type coordinate in the GEM model. The strategy has been to add the new coordinate option with minimal changes to the dynamical core and the rest of the GEM model source code. Therefore, the spatial grid structures in GEM-H, both in the horizontal and the vertical, are kept the same as those in GEM-P, which implies a staggered Arakawa C grid \citep{ara88} in the horizontal and a staggered Charney-Phillips grid \citep{cph53} in the vertical. The horizontal and vertical grid structures are presented in Fig. \ref{f_grid}. 

In addition to being similar to GEM-P for the limited-area model (LAM) grids, the global grid system is also kept unchanged in GEM-H, and is therefore, based on a Yin-Yang grid system (Qaddouri and Lee 2011). The Yin-Yang system combines two overlapping latitude-longitude LAM grids to form a global grid following the Schwarz method for non-matching domain decomposition \citep{qll08} and thus avoids pole-related singularity and convergence issues associated with a conventional global lat-lon grid. Further details on the Yin-Yang grid are provided by \cite{qle11}.\\

\subsection{Discretization in time}
The general form of an individual equation in the system comprised of Eqs. (\ref{e_u_1})--(\ref{e_t_1}) and (\ref{e_k_1}) can be expressed as
\begin{equation}
 \frac{dF_i}{dt}-G_i=P_i
 \label{e_gen_1}
\end{equation}
where $F_i$ denotes the advected quantity for an individual equation $i$ within the system, $G_i$ is the associated dynamics source term with linear and nonlinear components, and $P_i$ denotes the corresponding physics forcing. Similarly to GEM-P, treating the advection terms in a semi-Lagrangian way and applying a two-time-level Crank-Nicholson temporal discretization leads to
\begin{equation}
 \frac{F_i^A-F_i^D}{\Delta{t}}-\frac{1+b_i}{2}G_i^A-\frac{1-b_i}{2}G_i^D = \mathbf{s_c}\overline{P_i}
 \label{e_gen_2}
\end{equation}
where $\Delta t$ indicates the time-step length, and the superscripts $A$ and $D$ denote the arrival and departure positions of the air parcels at the current time $t$ and the previous time $(t-\Delta t)$, respectively. The integrals of the source terms $G_i$ for the different dynamical equations are approximated by trajectory averages. The parameter $b_i$ denotes the off-centering weight factor for the averaging of the dynamics source terms. When $b_i=0$, the averaging of the source term is fully centered, whereas $b_i>0$ implies additional weight placed on the implicit component of the source term. 

Historically, off-centering was implemented in GEM-P primarily to address spurious resonance originating from stationary orographic forcing \citep{rsr94}. However, it also suppresses computational noise and improves numerical stability. These other beneficial impacts have been found to be equally important in the current and previous implementations of GEM-P for the different operational NWP systems at ECCC. As the principal objective of this study is to have a GEM-H dynamical core that is equivalent to GEM-P for the different operational GEM-based NWP systems, off-centering has been retained in GEM-H. Also, following the latest implementation of GEM-P \citep{hgi17}, a differential approach for off-centering has been adopted for GEM-H where $b_i$ varies depending on the dynamical equation denoted by the subscript $i$. At present, the system of equations, given by Eqs. (\ref{e_u_1})--(\ref{e_t_1}) and (\ref{e_k_1}), are separated into three groups with an associated off-centering parameter for each group as follows:
\begin{enumerate}[(i)]
 \item {$b_m$ for the horizontal momentum equations [Eqs. (\ref{e_u_1})--(\ref{e_v_1})],}
 \item {$b_h$ for the continuity and thermodynamic equations [Eqs. (\ref{e_c_1})--(\ref{e_t_1})], and}
 \item {$b_{nh}$ for the vertical momentum and kinematic equations [Eqs. (\ref{e_w_1}) and (\ref{e_k_1})].}
\end{enumerate}

On the RHS of Eq. (\ref{e_gen_2}), the term $\overline{P_i}$ denotes the parameterized physics source term and the parameter $\mathbf{s_c}$ indicates the mode of coupling between physics and dynamics. Depending on the chosen method for dynamics-physics coupling, the value of $\mathbf{s_c}$ can be either 0 or 1. Also, the approach for dynamics-physics coupling determines how the contribution of $\overline{P_i}$ is accounted for in the model. Further discussions regarding the coupling of the parameterized physics forcing with the dynamical core is presented in section \ref{s_coupling}.

\subsection{Trajectory calculations}
Semi-Lagrangian treatment of advection requires the solution of kinematic displacement equations of the form
\begin{subequations}
  \begin{align}
    \dv{X}{t}=u,  \dv{Y}{t}=v, \\
    \dv{\zeta}{t}=\dot{\zeta},
  \end{align}
  \label{e_traj_1}
\end{subequations}
to determine the departure positions of the air parcels. In the context of GEM-P, \cite{hgi17} have shown that the consistency in the numerical discretizations between the dynamical and trajectory equations is fundamentally important for accurate solution of the advection problem. In order to be numerically consistent, similarly to the treatment of the dynamics source term in Eq. (\ref{e_gen_1}), the averaging of the velocities in Eq. (\ref{e_traj_1}) needs to be done using the trapezoidal rule. Furthermore, the interpolation scheme employed to determine the wind field at the departure positions for the trajectory calculations need to be the same as the one applied to determine the source terms in the dynamical equations at the departure positions. In the case of GEM-P, cubic interpolation is used for both the wind field and the dynamical source terms to achieve numerical consistency. Following the conclusions of \cite{hgi17}, similar consistent trajectory calculation approach is adopted in GEM-H, i.e., trapezoidal rule for evaluating the integral of the source term in Eq. (\ref{e_traj_1}) along with cubic interpolation to determine the wind field at the departure positions. 

\subsection{The elliptic problem}
In order to solve the system of equations associated with the GEM-H formulation, each equation of the form (\ref{e_gen_2}) is rearranged to separate the linear and nonlinear components of the implicit part and is expressed as
\begin{equation}
 L_i=R_i-N_i,
 \label{e_lnr}
\end{equation}
where $L_i=(F^A_i/\tau_i-G^A_i)_{linear}$, $N_i=F^A_i/\tau-G^A_i-L_i$, and $R_i=F^D_i/\tau_i+\beta_iG^D_i+\mathbf{s_c}\overline{P_i}$ with $\beta_i=(1-b_i)/(1+b_i)$ and $\tau_i=\Delta t(1+b_i)/2$. Eqs. (\ref{e_u_1})--(\ref{e_t_1}) and (\ref{e_k_1}) are rearranged as in Eq. (\ref{e_lnr}), giving
\begin{equation}
 L_u = \frac{u}{\tau_m} + \delta_Xq - \mathbf{s_i } J_X\overline{J^{-1}_\zeta\delta_\zeta q}^{X\zeta},
 \label{e_l_u}
\end{equation}
\begin{equation}
L_v = \frac{v}{\tau_m} +\delta_Yq - \mathbf{s_i } J_Y\overline{J^{-1}_\zeta\delta_\zeta q}^{Y\zeta},
 \label{e_l_v}
\end{equation}
\begin{equation}
L_w = \frac{w}{\tau_{nh}} + (\mathbf{s_i}J^{-1}_\zeta + \mathbf{s_d})\delta_\zeta q - g\frac{T^{'}_v}{T_v},
 \label{e_l_w}
\end{equation}
\begin{equation}
 L_c=\frac{1}{\tau_h}\bigg(\frac{q}{c_*^2}+\ln{J_\zeta}\bigg)+\delta_Xu+\frac{1}{\cos_\phi}\delta_Y(\cos\phi_v)+\delta_\zeta\dot{\zeta}-\varepsilon\overline{w}^{\zeta},
 \label{e_l_c}
\end{equation}
\begin{equation}
 L_T=\frac{1}{\tau_h}\bigg(\frac{T^{'}_v}{T_v}-\frac{\overline{q}^\zeta}{c_{pd}T_*}\bigg)+\mu w,
 \label{e_l_t}
\end{equation}
\begin{equation}
 L_z=\frac{z-\zeta}{\tau_{nh}}+\dot{\zeta}-w,
 \label{e_l_k}
\end{equation}
where the symbol $\delta_i$ denotes the finite difference operator along the $i$-direction, and the overline operator $\overline{( )}^j$ implies spatial averaging in the $j$-coordinate. For convenience of notation the terms $\frac{g}{c^{2}_{*}}$ and $\frac{N^{2}_{*}}{g}$ have been replaced by $\varepsilon$ and $\mu$, respectively. The corresponding nonlinear components $N_i$ associated with the discretized forms of the Eqs. (\ref{e_u_1})--(\ref{e_t_1}) and (\ref{e_k_1}) as well as the associated RHS terms $R_i$ are provided in Appendix A. The parameters $\mathbf{s_i}$ and $\mathbf{s_d}$ in the above equations denote the choice of the solver for the dynamical core, where the subscripts $i$ and $d$ stand for $iterative$ and $direct$ respectively. Based on the selected solution approach, these parameters can be either 1 or 0, and are mutually exclusive such that $\mathbf{s_i}=1-\mathbf{s_d}$. Further discussion on the solution approaches is presented in Section \ref{s_solver}. As shown by \cite{cst88}, the solution of the system of equations of the type (\ref{e_lnr}) requires nonlinear iterations for convergence, where the nonlinear terms $N_i$ are re-evaluated during each iteration using the latest values of the prognostic variables. Furthermore, Crank-Nicholson iterations are required, where the $R_i$ terms are re-evaluated during each iteration at the departure positions calculated using the latest velocity estimates. At present, the GEM-based operational NWP systems at ECCC utilize two Crank-Nicholson iterations and within each Crank-Nicholson step two nonlinear iterations are carried out. As a result, irrespective of the solution approach, the solver is called four times during each dynamical time step.

The discretized equations with the left hand sides (LHSs) given by Eqs. (\ref{e_l_u})--(\ref{e_l_k}) are then reduced into a single elliptic boundary value (EBV) problem through elimination of variables, where the LHS of the final elliptic problem has the form
\begin{equation}
 L^{'''}_c=\delta_XA+\frac{1}{\cos\phi}\delta_Y[\cos\phi B]+\delta_\zeta C-\varepsilon\overline{C}^\zeta-\gamma q,
 \label{e_l_ebv}
\end{equation}
where $A=\bigg(\delta_Xq - \mathbf{s_i } J_X\overline{J^{-1}_\zeta\delta_\zeta q}^{X\zeta}\bigg)$, $B=\bigg(\delta_Yq - \mathbf{s_i } J_Y\overline{J^{-1}_\zeta\delta_\zeta q}^{Y\zeta}\bigg)$ and $C=\Gamma\bigg[(\mathbf{s_i}J^{-1}_\zeta+\mathbf{s_d})\delta_\zeta q-\mu\overline{q}^\zeta\bigg]$. Also, in Eq. (\ref{e_l_ebv}), $\gamma=1/(c^2_*\tau_h\tau_m)$ and $\Gamma=1/(\tau_m/\tau_{nh}+N^2_*\tau_h\tau_m)$. It is important to note that, with $\mathbf{s_i}=1$, the terms $A$, $B$ and $C$ (with $\mu=0$) are simply the components of the gradients in the terrain-following coordinate system. The sequence of steps involved in deriving the EBV problem, i.e., the final form of $L^{'''}_c$, is provided in Appendix B.

\subsection{Initial and boundary conditions}
As with the case of any initial value problem, in order to initiate integration in time, the GEM-H dynamical core requires initial values of all the prognostic variables. At present, ECCC's operational data assimilation system provides analyzed initial values for the horizontal wind components $u$ and $v$, virtual temperature $T_v$ and surface pressure $p_s$. The remaining prognostic variables  $w$, $\dot{\zeta}$ and $q$ are computed in a diagnostic manner at time $t=0$. The initial value of $q$ is obtained from the analyzed surface pressure $p_s$ with a hydrostatic approximation. Substituting $\dv{w}{t}=0$ in Eq. (\ref{e_w_1}) gives
\begin{equation}
\frac{1}{J_{\zeta}}\pdv{q}{\zeta}=g\frac{{T_v}^{'}}{T_v}
 \label{e_hydro}
\end{equation}
as a hydrostatic approximation. The value of $q$ at the different model levels are then obtained by integrating  Eq. (\ref{e_hydro}) where at the surface, due to the hydrostatic approximation, $q_s=R_dT_*\ln (p_s/p_*)$. The initial value of $\dot{\zeta}$ is computed by assuming $\pdv{\rho}{t}=0$ at time $t=0$ in the continuity equation. In the $\zeta$-coordinate, this takes the form
\begin{equation}
 \pdv{}{X}\bigg(\rho u \pdv{z}{\zeta}\bigg)+\frac{1}{\cos\phi}\pdv{}{Y}\bigg(\cos\phi\rho v \pdv{z}{\zeta}\bigg)+\pdv{}{\zeta}\bigg(\rho \dot{\zeta} \pdv{z}{\zeta}\bigg)=0,
 \label{e_zdot_ini}
\end{equation}
which is then discretized to compute the initial value of $\dot{\zeta}$. Once $\dot{\zeta}$ is known, the initial value of $w$ is obtained from its definition in the $\zeta$-coordinate. i.e., $w\equiv\dv{z}{t}=uJ_X+vJ_Y+\dot{\zeta}J_\zeta$. 

The boundary conditions for the upper and lower boundaries are given by $\dot{\zeta}_T=\dot{\zeta}_S=0$. This implies that the vertical motion in the $\zeta$-coordinate vanishes at the surface and the model top which is flat. For LAM problems, GEM-H also requires lateral boundary conditions which are obtained from the driving fields. As the global Yin-Yang system is  based on two interacting geometrically identical LAM domains, it therefore similarly requires lateral boundary conditions. In this case, the boundary conditions for one sub-domain (Yin or Yang) depend on the solution in the other. Thus the solution of the global problem is obtained by iteratively solving the two sub-problems separately and updating the values in the overlapping region until a certain convergence criteria is satisfied \citep{qle11}.

\section{Solution of the EBV problem} \label{s_solver}
The EBV problem to be resolved by GEM-H at each model time-step can be expressed as
\begin{equation}
\nabla_\zeta^2q + \mathbf{M}q=\mathbf{R},
 \label{e_ebv_gf}
\end{equation}
by replacing $L^{'''}_c$ in Eq. (\ref{e_l_ebv})  with $(\nabla^2_\zeta+\mathbf{M})q$, where $\nabla_\zeta^2=(\delta_{XX} + \frac{1}{\cos \phi}\delta_Y (\cos\phi \delta_Y))$ is a discretized two-dimensional horizontal operator in the $\zeta$-coordinate, $\mathbf{M}$ contains all the remaining terms of $L^{'''}_c$ that include the discretized difference and averaging operators in the horizontal and vertical dimensions, $q$ is the unknown and $\mathbf{R}$ includes the explicit RHS terms as well as the implicit nonlinear terms. It is important to note that the nonlinear terms in $\mathbf{R}$ require iterations for convergence, irrespective of the solution approach. At present, the GEM dynamical core uses two iterations for the sufficient convergence of the nonlinear terms and two iterations for trajectories. As a result, the solver is called into action four times during each dynamical step.

Once Eq. (\ref{e_ebv_gf}) is solved to obtain the unknown $q$, the other prognostic variables are obtained through back substitution as presented in Appendix C. As has been mentioned earlier, two general approaches are available for solving the elliptic problem -- direct and iterative. The selection of these approaches depends on which terms are included in $\mathbf{M}$, and is determined by the values of the terms $\mathbf{s_i}$ and $\mathbf{s_d}$ in Eqs. (\ref{e_l_u})--(\ref{e_l_t}).

\subsection{The direct solver}
The direct solver works by first decoupling Eq. (\ref{e_ebv_gf}) in the vertical. It is achieved through the expansion of the unknown $q$ and the RHS $\mathbf{R}$ in terms of the eigenvectors that diagonalizes the operator $\mathbf{M}$ \citep{qle11}. This is only possible when the operator $\mathbf{M}$ does not include contributions from the metric terms arising from the vertical coordinate transformation that involves horizontally variable coefficients imparting horizontal coupling. Therefore, the implementation of direct solver in GEM-H requires a `simplified approach' where all the metric terms of the relevant discretized equations are treated as nonlinear terms. This is achieved by setting $\mathbf{s_d}=1$. For $N_k$ number of vertical levels used in the model, vertical separation reduces Eq. (\ref{e_ebv_gf}) to a set of $N_k$ independent horizontal Helmholtz problems of the form
\begin{equation}
\nabla_\zeta^2\tilde{q} + m\tilde{q}=\tilde{\mathbf{R}},
 \label{e_helmh}
\end{equation}
where $\tilde{q}$ and $\tilde{\mathbf{R}}$ are the vertical projections of $q$ and $\mathbf{R}$, respectively, and $m$ is the eigenvalue of the operator $\mathbf{M}$. The horizontal solution of the algorithm then proceeds by expanding $\tilde{q}$ and $\tilde{\mathbf{R}}$ in terms of the eigenvectors that diagonalize the $X$-component of the two-dimensional operator $\nabla_\zeta^2$. For $N_i$ number of grid points along the longitude $X$, this leads to $N_i$ independent tridiagonal problems of $N_j$ dimension for each model vertical level, where $N_j$ denotes the number of grid points along the latitude $Y$. The total number of tridiagonal problems to solve is therefore $N_k\times N_i$. Solution to the tridiagonal problems are computed by Gaussian elimination without pivoting, and afterwards, the final three-dimensional solution $q$ is reconstituted \citep{qle10}.

The `simplified approach' has been primarily implemented in GEM-H to take advantage of the computational performance of the direct solver. The direct solver uses Fast Fourier Transform to compute the horizontal solution $\tilde{q}$ and outperforms any iterative approach implemented at the ECCC by a substantial margin for the configurations of the various GEM-based NWP systems running operationally. The `simplified approach' thus makes the application of GEM-H for such configurations feasible and keeps the option of a future replacement of the GEM-P core with GEM-H. The simplified approach, however, works as long as the vertical-horizontal coupling, imparted through the metric terms, is not too significant so that these terms can be treated efficiently through nonlinear iterations. This approach works unless the maximum terrain slope is not sufficiently steep (less than 30$^\circ$) which is generally the case for most of the operational GEM-based NWP systems. However, with increasing spatial resolution, the slopes in grid-scale orography also increase, particularly over complex terrain, which leads to increased vertical-horizontal coupling and, at one point, makes the `simplified approach' inapplicable. 

\subsection{The iterative solver}
When all the metric terms in $A$ and $B$ (see Eq. \ref{e_l_ebv}) are included in $\mathbf{M}$, the resulting vertical-horizontal coupling makes the problem non-separable. This is done by setting $\mathbf{s_i}=1$. In such a scenario, the three-dimensional equation of the form (\ref{e_ebv_gf}) can only be solved at each time step by using an iterative solver. The current iterative solver for the EBV problem in GEM-H is based on the flexible generalized minimal residual (FGMRES) method \citep{saa93, qle10}. 

The fully discretized system of equations of the form (\ref{e_ebv_gf}) can be further generalized as
\begin{equation}
 \mathbf{A}q=\mathbf{R}
 \label{e_axb}
\end{equation}
where coefficient matrix $\mathbf{A}$ contains the discretized operator $(\nabla_\zeta^2 + \mathbf{M})$. The FGMRES method approximates the solution in a Krylov sub-space of small dimension by minimizing the Euclidean norm of its residual. A major advantage of such a method is that instead of explicitly generating the coefficient matrix  $\mathbf{A}$, one only needs to compute the vector resulting from the action of the underlying operator $(\nabla_\zeta^2 + \mathbf{M})$ on the vector $q$. Efficient functioning of such an iterative solver, however, requires a pre-conditioner. At present, the pre-conditioner is based on the block Jacobi iteration for the EBV in Eq. (\ref{e_ebv_gf}), where all the metric terms in $\mathbf{M}$ are absent. This pre-conditioner improves the convergence rate of the FGMRES solver. However, the time of execution is still high compared to the fast direct solver. Significant research is currently underway at ECCC to devise iterative solvers that are competitive with the direct approach and will work for both GEM-P and GEM-H. At present, the current implementation of the iterative solver in GEM-H---although not as efficient---provides the necessary reference for the direct solver approach.

\section{Dynamics-physics coupling} \label{s_coupling}
Along with the dynamical core, parameterization of the subgrid-scale physical processes constitutes the other fundamental component of any NWP model. Coupling between the dynamical core and the parameterized subgrid-scale physical processes is of critical importance. How to devise the most appropriate coupling strategy is still an unsettled question \citep{bbb18}. This issue is being actively studied at ECCC. However, it is not the objective of this study to delve deep into the fundamental questions around dynamics-physics coupling. Rather, in this section, the issue of coupling the RPN Physics Package with the GEM dynamical core is discussed in order to determine which approach is the most feasible for GEM-H among the options that are available for GEM-P. 

It is important to note that, during every model time step in GEM, the RPN Physics Package is executed after the dynamical equations have been resolved by the dynamical core and thus the physics schemes utilize the outputs of the dynamics step as inputs. However, irrespective of the vertical coordinate used in the dynamical core, the physics schemes use a traditional $\sigma$-coordinate in the vertical, defined as 
\begin{equation}
 \sigma=\frac{\pi}{\pi_s},
 \label{e_sigma}
\end{equation}
where $\pi$ is the hydrostatic pressure. Also, the physics schemes work within a one-dimensional configuration where each processor core only has access to the vertical structure of the meteorological fields associated with a single horizontal grid point. The various physical processes are parameterized sequentially where the tendencies estimated by one parameterization scheme affects the ones that follow. The parameterization sequence during each physics step initiates with the radiation scheme and is followed by the parameterizations of the surface processes, gravity wave drag, boundary layer turbulence, convection and grid-scale condensation. At the end of the physics step, the tendencies from the different physical parameterization schemes are aggregated to compute the grid-scale tendencies for the wind components, temperature, water vapor and the hydrometeors.

\subsection{Different coupling approaches in GEM}
Particularly in the context of GEM-P, two approaches are presently available to couple the RPN Physics Package with the GEM dynamical core. A brief discussion on these methods will be helpful in establishing the rationale behind the approach selected for the GEM-H dynamical core.

\begin{enumerate}[(i)]
 \item {Split method: As has been mentioned earlier, $\mathbf{s_c}$ determines the mode of coupling between dynamics and physics. If $\mathbf{s_c}=0$ then the dynamical equations are resolved in the absence of any physical forcing and at the end of the dynamics step their contributions are incorporated as adjustments in the so called `split mode'. In the absence of physics forcing, Eq. (\ref{e_gen_2}) becomes
\begin{equation}
 \frac{F_i^{A*}-F_i^D}{\Delta{t}}-\frac{1+b_i}{2}G_i^{A*}-\frac{1-b_i}{2}G_i^D = 0
 \label{e_gen_3}
\end{equation}
where $F_i^{A*}$ is the interim solution of the dynamics step. Once Eq. (\ref{e_gen_3}) is resolved, the physics source term is then applied as grid point adjustments as follows
\begin{equation}
 (\delta F)_{phys}=F_i^{A} - F_i^{A*} = \Delta t P_i.
 \label{e_gen_4}
\end{equation}
Thus, in the split method, the dynamics step predicts an inviscid and adiabatic solution that is modified through adjustments attributable to the parameterized physics forcings in order to obtain the complete solution at the end of each model time step. 
}

 \item {Explicit method: The second option for dynamics-physics coupling treats the physics source terms explicitly by setting $\mathbf{s_c}=0$ and replacing $\overline{P_i}$ by $P_i^D$. This method is referred to as the `explicit method' and moves by solving equations of the following form at each model time step
\begin{equation}
  \frac{F_i^A-F_i^D}{\Delta{t}}-\frac{1+b_i}{2}G_i^A-\frac{1-b_i}{2}G_i^D = P_i^D.
 \label{e_gen_7}
\end{equation}
In this approach, physics tendency $P_i$ from the previous time step is combined with the RHS term $R_i$, followed by the determination of $R_i^D$ at the departure positions in a semi-Lagrangian way. In other words, the explicit method works by directly incorporating the impact of physics forcings as tendencies into the discretized dynamical equations.}
 
\end{enumerate}

\subsection{Coupling in GEM-H}
Although both of the aforementioned coupling methods are available for GEM-P, it is important to note that all the operational NWP systems based on GEM-P at present utilize the split method for dynamics-physics coupling. Nevertheless, there exists strong concern about the split method in general, and a brief discussion highlighting the pertinent issues will be helpful. 

In the context of model formulations for both GEM-P and GEM-H, the term $F_i$ does not necessarily coincide with the prognostic model variables. For example, in the case of GEM-P in its hydrostatic mode, as presented by \cite{gdm14}:
\begin{subequations}
  \begin{align}
    F=\left\{u,v,Bs+\ln(1+\pdv{B}{\zeta}),\ln\frac{T_v}{T_*}-\kappa_dBs\right\}, \\
    P=\left\{\dv{}{t}\bigg(u,v,\ln\rho,\ln T_v\bigg)\right\}_{phys},
  \end{align}
  \label{e_couple_gemp}
\end{subequations}
whereas the prognostic variables are $u$, $v$, $s$, $\dot{\zeta}$, and $T_v$. Therefore, in the actual model implementation of the split method, the prediction from the dynamics step is utilized to compute the interim state of the prognostic variables and adjustments are then applied to these variables at the end of the physics step. It is important to note that in this case, the only adjustments that have been found to not result in any issue of major concern are $(\delta u)_{phys}$, $(\delta v)_{phys}$ and $(\delta T_v)_{phys}$. Also in GEM-P, adjustments are required for density, as in effect
\begin{equation}
 \left(\dv{\ln \rho}{t}\right)_{phys}\equiv\dv{}{t}\ln(1+r_w).
 \label{e_split_density}
\end{equation}
Although water vapor and hydrometeors are updated through physical parameterizations, the only variable that could be updated in the split mode appears to be $s$ through a surface pressure adjustment $\delta\pi_s$ that may be computed as
\begin{equation}
 \delta\pi_s=\int{\delta \ln(1+r_w)d\pi},
 \label{e_delta_pis}
\end{equation}
which takes into account the net inflow/outflow of mass through the Earth's surface at every model grid point. Unfortunately, the vertical distribution of this change in mass through water vapor and precipitation fluxes cannot be correctly accounted for in the split mode, and is found to produce considerable noise in the wind forecast.

\begin{figure}
\noindent\includegraphics[width=45pc,angle=0]{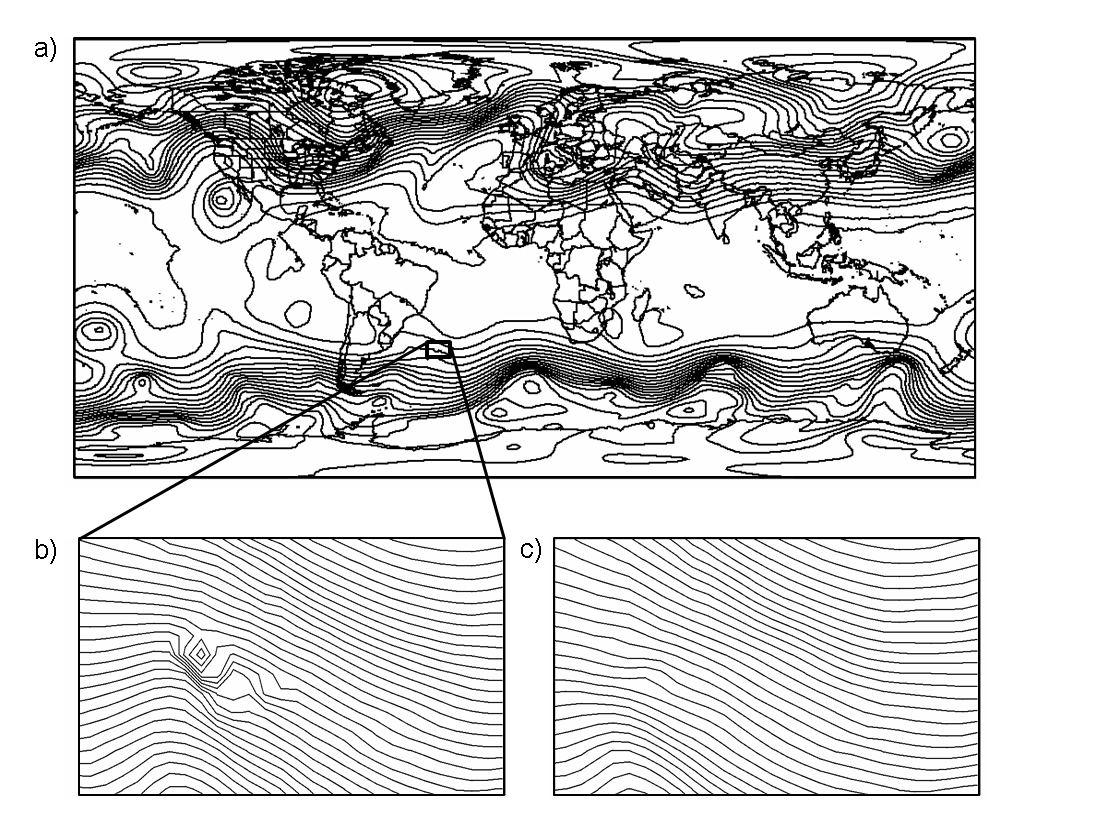}\\
\caption{a) Geopotential height (dam) at 400 hPa after 72 hours for a 25-km global forecast initiated at 1200 UTC of 25 January 2015 obtained with GEM-P using the `split method' for dynamics-physics coupling (contour intervals of 6 dam). b) The enlarged view over a small section of the global domain (contour intervals of 0.5 dam). c) Same as in Fig. \ref{f_split}b, but with `explicit method' for the dynamics-physics coupling.}
\label{f_split}
\end{figure}

Furthermore, in the context of three-time-level discretization, \cite{clz98} have shown that the split method can lead to erroneous results for long time steps that are permissible with the semi-implicit semi-Lagrangian models. Similar conclusions were drawn from a theoretical analysis for two-time-level schemes by \cite{swc02}. Figure \ref{f_split}a shows the geopotential height contours at 400 hPa from a 72-h global forecast with ECCC's 25-km resolution Global Deterministic Prediction System (GDPS) using the GEM-P dynamical core. The results correspond to split method for dynamics-physics coupling. Although the distribution of geopotential height for the meso and large scales does not reveal any issue of immediate concern, when one looks at the smallest scales, i.e., scales of about a few grid lengths, some spurious computational noise is visible (see Fig. \ref{f_split}b). Historically, the operational NWP systems at ECCC have been utilizing spatial filters over the model-predicted meteorological fields of interest, like the geopotential height, to smooth out any computational noise in the model outputs. As a result, the kind of computational noise shown in Fig. (\ref{f_split}b) has not been troublesome for the meteorologists using ECCC's operational NWP outputs. Nevertheless, the computational noise associated with the split method remains as a concern. However, as shown in Fig. \ref{f_split}c, when the split method for coupling is replaced by the explicit method, the noise in the geopotential height disappears. It should be noted that, although explicit coupling can impose stability limitations in terms of the acceptable length of time steps, ECCC's operational NWP system configurations are found to function with explicit coupling without requiring any adjustments to the time-step lengths.

Similarly to GEM-P, the $F_i$ terms do not coincide with the prognostic variables for all of the GEM-H model equations, where
\begin{subequations}
  \begin{align}
    F=\left\{u,v,w,\frac{q}{c^2_*}+\ln J_\zeta,\ln\frac{T_v}{T_*}-\frac{q}{c_{pd}T_*}\right\}, \\
    P=\left\{\dv{}{t}\bigg(u,v,w,\ln\rho T_v,\ln T_v\bigg)\right\}_{phys}.
  \end{align}
  \label{e_couple_gemh}
\end{subequations}
Particularly, the presence of the physics forcing term $\left(\dv{\ln \rho T_v}{t}\right)_{phys}$ in the modified continuity equation (\ref{e_c_1}) poses an additional challenge for the GEM-H formulation, as far as the split method is concerned. If one attempts to apply this tendency as a hydrostatic adjustment to pressure in the split mode then of course its effect is not limited to the continuity equation alone. Rather, applying the adjustments to pressure, i.e. changing $q$ in the case of GEM-H, also affects the thermodynamic equation. Such an adjustment leads to over compensation and is found to result in spurious bias in the temperature in upper troposphere and stratosphere, which is unacceptable (not shown). Also, the jet-level wind is found to be adversely affected. As a result, in GEM-H the physics contributions are accommodated through the explicit method by including them as explicit grid-scale tendencies in the RHS of the discretized dynamical equations. For a fair comparison between the new and the existing dynamical core, the results for both the GEM-P and GEM-H cores presented in the rest of this paper are obtained with explicit dynamics-physics coupling. It is also important to mention that, even though parameterization of physical processes like boundary-layer turbulence can modify vertical motion $w$, at present the impact of physics on $w$ is neglected. This is the case for both GEM-P and GEM-H.

A hybrid `split-explicit method' has also been tested with GEM-H, where the physics contributions to the thermodynamic and horizontal momentum equations are accounted for through the `split method' while the contributions to the continuity equation is accommodated using the `explicit method'. Such a hybrid approach with GEM-H produces results that are equivalent to those obtained with the `split method' for GEM-P. However, questions remain about the numerical consistency of the hybrid split-explicit method.

It is important to mention that the current implementation of the RPN Physics Package, when coupled with the GEM-P (GEM-H) dynamical core through the explicit method, leads to some deterioration in temperature bias in the upper troposphere compared to the split (split-explicit) method. This implies that further research is necessary to have a more consistent dynamics-physics coupling with the explicit method. Particularly, the grid-scale condensation scheme in GEM, for 10 km or coarser horizontal resolutions, has been found to exhibit large sensitivity with the explicit method which leads to under-prediction of clouds (R. McTaggart-Cowan, ECCC, personal communication). This implies that some process-specific adjustments in the computation of the relevant physics tendencies may be required to improve the overall dynamics-physics coupling. The challenges imposed by the process-specific issues are also being explored by other operational NWP centers \citep{bbb18}. Currently, work is underway at ECCC to explore the various issues within the coupling interface as well as the parameterizations of the different physical processes to improve the dynamics-physics coupling in general. Further discussion on this issue, however, is beyond the scope of this paper.

\section{Evaluation of GEM-H} \label{s_evaluation}
One of the most important objectives of this study has been to develop a dynamical core for the GEM model that  utilizes a height-based TFC and is capable of producing predictions that are equivalent to the results obtained by the pressure-based dynamical core.  In order to evaluate the consistency and performance of the new dynamical core, a number of numerical experiments covering a wide range of scales---ranging from microscales to the meso and synoptic scales---have been carried out. These include two-dimensional theoretical test cases involving bubble convection \citep{rob93} and nonhydrostatic mountain waves \citep{slf02} as well as three-dimensional global NWP. The two-dimensional test cases selected in this paper have become ubiquitous tools in testing the consistency and performance of nonhydrostatic dynamical cores. Also, the availability of the GEM-P core provides the opportunity to have reference solutions for all these cases.

\subsection{Robert's bubble convection case}
\cite{rob93} presented a two-dimensional theoretical case involving the evolution of a warm bubble within a dry isentropic atmosphere. Initially, the bubble has a diameter of 500 m and is placed 10 m above a flat surface within a 1 km $\times$ 1 km computational domain, and has a uniform potential temperature of 30.5$^\circ$C. Also, the basic-state atmosphere is at rest under a hydrostatic equilibrium with an isentropic basic-state temperature of 30$^\circ$C. As the bubble has a potential temperature excess of 0.5$^\circ$C compared to the surrounding atmosphere, it rises due to the action of the buoyancy force. The absence of any orographic variation makes this experiment an excellent benchmark to test the functioning of advection and buoyancy within a dynamical core during the early stages of its development.

\begin{figure}
\noindent\includegraphics[width=45pc,angle=0]{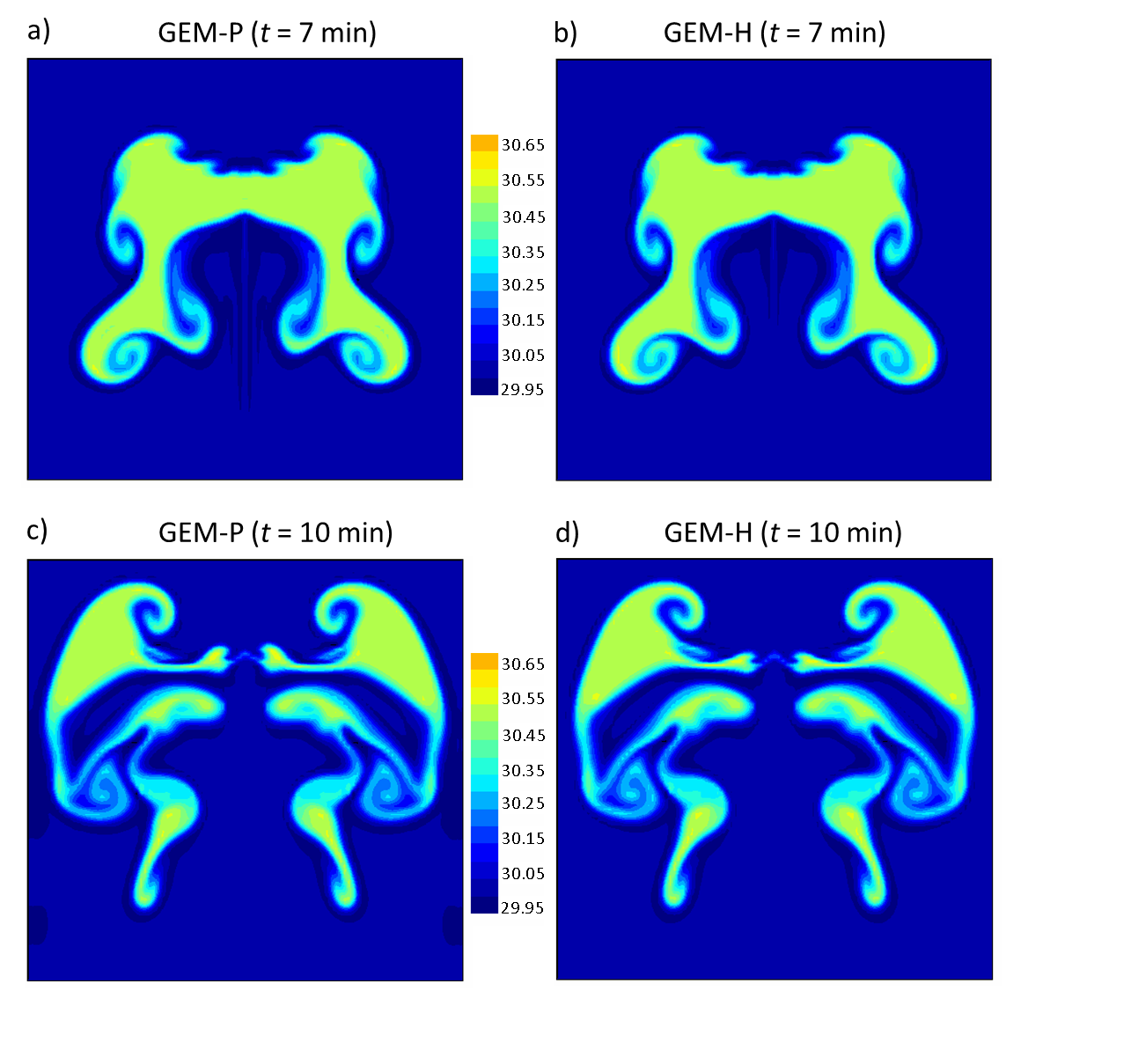}\\
\caption{Potential temperature ($^\circ$C) distribution for the rising bubble with: a) GEM-P at $t=7$ min, b) GEM-H at $t=7$ min, c) GEM-P at $t=10$ min, and d) GEM-H at $t=10$ min.}
\label{f_bubble}
\end{figure}

The numerical experiment for bubble convection is carried out with a spatial grid resolution of 10 m and a time step of 5 s. No explicit numerical diffusion is used. The resulting evolution of the bubble, in terms of its potential temperature distribution at two different times (7 min and 10 min), is presented in Fig. \ref{f_bubble} for both GEM-P and GEM-H. The bubbles predicted by the two cores initially deform into a somewhat mushroom-like shape (see at $t=7$ min) and then are deformed further (at $t=10$ min). Overall, the predictions from GEM-P and GEM-H are equivalent for the entire range of scales - from the large to the smallest scales. Such a good resemblance between the two predictions imply negligible impact of the choice of the vertical coordinate and the other modifications in model formulations in the absence of any orographic variation at the model surface. It also indicates that the representation of the advection and buoyancy effects are comparable between the two GEM cores. It is important to note that, due to considerable differences in the model formulations and spatiotemporal discretizations, it is difficult to compare the evolution of the bubbles between two completely separate models in a quantitative manner. Only qualitative comparisons are feasible. Therefore, the lesser resemblance between the results from \cite{rob93} and the GEM dynamical cores are not unusual. Although the predictions from the two GEM cores have some large-scale resemblance to the results presented by \cite{rob93}, significant differences appear at the upper half of the bubble - particularly at $t=10$ min. However, the upper structure of the bubble compares better with the predictions by \cite{spu92}. Also, because of the implicit dissipation associated with the semi-Lagrangian approach, the GEM solution does not suffer from computational noise like models based on Eulerian advection \citep{jua00}. Overall, as GEM-P is being used operationally at ECCC, the resemblance between GEM-H and GEM-P is of more significant importance, as it confirms consistency of the GEM-H formulation and a neutral impact of the vertical coordinate modification on buoyancy and advection.

\subsection{Sch\"{a}r's mountain case}
\cite{slf02} presented a linear two-dimensional theoretical test case of mountain waves which is an excellent benchmark for verifying nonhydrostatic dynamical cores, particularly in determining the presence of possible inconsistencies in the numerical details \citep{hgi17, mdw10, gbd05, ksf03}. The bottom boundary profile of the idealized mountain for this case is defined by
\begin{equation}
z_s=z_0 e^{{-(x/a)}^2}\cos^2(\pi x/l_x),
 \label{e_schaer_mtn}
\end{equation}
where $z_0$=250 m, $l_x$=4 km, $a$=5 km, and $\pi$ is the conventional mathematical constant. The upstream flow conditions are given by uniform upstream wind $U$=10 m s$^{-1}$, upstream surface temperature $T_{surf}$=288 K, upstream surface pressure $p_0$=1000 hPa, and a constant Brunt-V\"{a}is\"{a}l\"{a} frequency $N_*$=0.01 s$^{-1}$. All other conditions for the simulations of this test case is similar to those presented by \cite{hgi17}.

\begin{figure}
\noindent\includegraphics[width=43pc,angle=0]{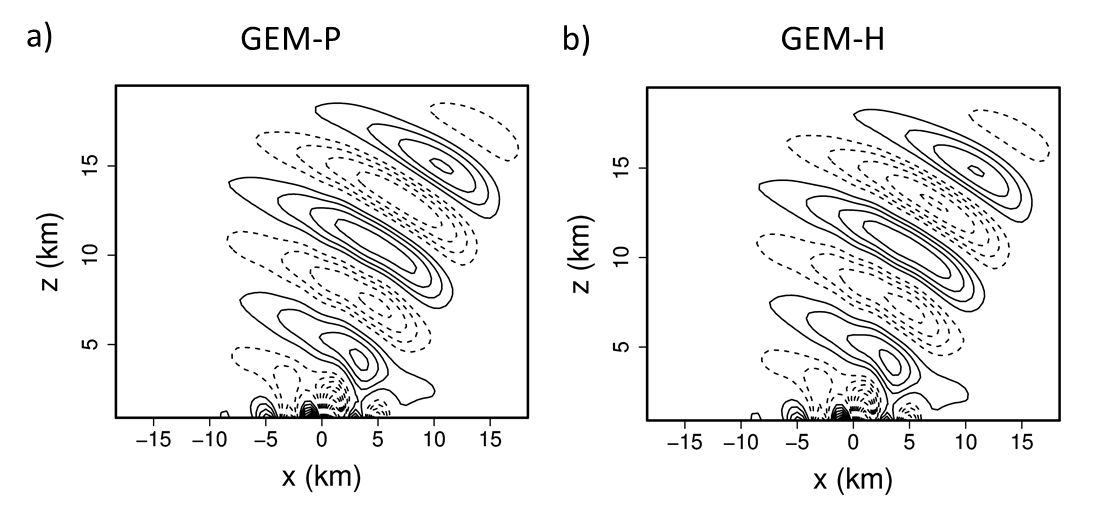}\\
\caption{Steady-state vertical velocity contours (at 0.1 m s$^{-1}$ intervals) for the Sch\"{a}r mountain case predicted by : a) GEM-P and b) GEM-H. No off-centering is used in the discretized dynamical and trajectory equations.}
\label{f_schaer_1}
\end{figure}

\begin{figure}
\noindent\includegraphics[width=43pc,angle=0]{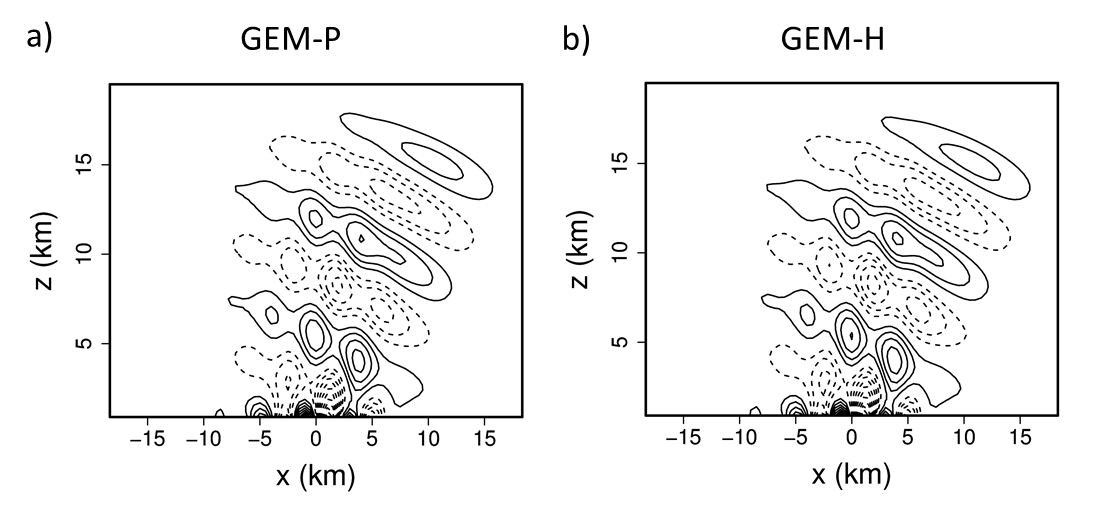}\\
\caption{Same as in Fig. \ref{f_schaer_1}, but with uniform off-centering ($b_m=b_h=b_{nh}=0.2$) used in the discretized dynamical equations. The absence of off-centering in the discretized trajectory equations leads to inconsistent trajectory calculations and distortions in the mountain waves.}
\label{f_schaer_2}
\end{figure}

\begin{figure}
\noindent\includegraphics[width=43pc,angle=0]{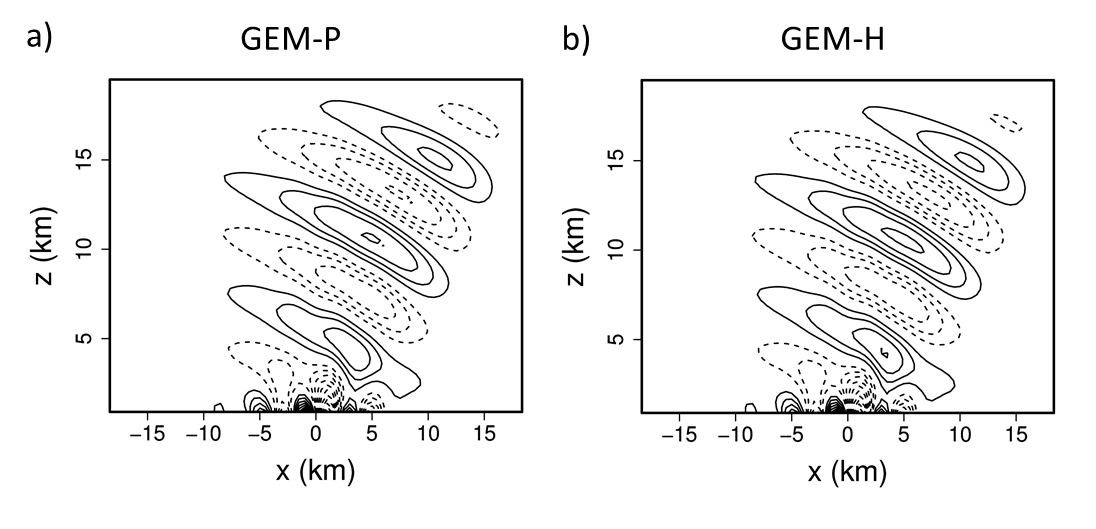}\\
\caption{Same as in Fig. \ref{f_schaer_1}, but with consistent off-centering, i.e., off-centering is applied to both the discretized dynamical and trajectory equations.}
\label{f_schaer_3}
\end{figure}

One major advantage of Sch\"{a}r's mountain case is the availability of a steady-state analytical solution of the corresponding to the linearized problem as a reference \citep{slf02}. The simulated quasi-steady  vertical velocity obtained after 4 hours of integration with both the GEM-P and GEM-H dynamical cores are presented in Fig. \ref{f_schaer_1}. The analytical solution of the problem, as presented by \cite{slf02}, is a combination of rapidly decaying small-scale nonhydrostatic mountain waves close to the surface and large-scale hydrostatic waves extending to much higher altitudes. As can be seen in Fig. \ref{f_schaer_1}, the solutions with both types of vertical coordinates generate these two regimes of the mountain waves and are very similar. Results shown in this figure represent simulations that have been carried out without any off-centering in the discretized dynamical equations and with consistent trajectory calculations, i.e., integration of the wind field in the trajectory equations is based on the trapezoidal rule while the wind field at the departure positions is obtained with cubic interpolation. Inconsistent trajectory calculations were found to produce similar distortion in the large-scale hydrostatic waves with GEM-H (not shown) as has been found for GEM-P earlier by \cite{hgi17}. Although the GEM-H results presented here corresponds to the direct solver based on the simplified approach, results with the iterative solver is found to be almost identical (not shown).
Figure \ref{f_schaer_2} reveals that off-centering in the discretized dynamical equations leads to distortions in the vertical velocity distribution, irrespective of the type of the vertical coordinate. The solutions presented here are obtained with uniform off-centering involving $b_m=b_h=b_{nh}=0.2$, which are the standard values used in the current GEM-based operational NWP systems. The results correspond to $\Delta t$=32 s, for which the maximum Courant number is approximately 0.76. Reducing the time step to even 4 s is unable to remove these distortions. Also, reducing the level of off-centering is found to reduce the level of distortion in the mountain waves, but the distortions are only completely eliminated when $b_m=b_h=b_{nh}=0$ (not shown). This conforms to the conclusions drawn by \cite{hgi17} in the context of GEM-P. Furthermore, as has been shown by \cite{hgi17}, consistent trajectory calculations in the presence of off-centering necessitates off-centered averaging applied to the integrals of the source term on the RHS of Eq. (\ref{e_traj_1}). With uniform off-centering applied to the discretized dynamical equations, the discretized trajectory equations also require uniform off-centering of the same degree. Figure \ref{f_schaer_3} reveals that in the presence of consistent off-centering in the trajectory calculations, the distortions in the vertical velocity distribution are eliminated for the both dynamical cores. In the presence of differential off-centering, i.e., with different values of $b_h$ and $b_{nh}$ for hydrostatic and nonhydrostatic contributions in the system of dynamical equations, a similar differential approach is required for off-centering in the discretized trajectory equations. As has been shown for GEM-P by \cite{hgi17}, in order to achieve numerical consistency, the off-centering in the discretized source terms in Eq. (\ref{e_traj_1}a) and (\ref{e_traj_1}b) need to be equal to the values of $b_h$ and $b_{nh}$ used in the discretized dynamical equations, respectively.

\subsection{Global deterministic prediction}
A series of 5-day global forecasts, with 25 km horizontal grid spacing, has been carried out covering winter and summer periods for the Northern Hemisphere to compare the predictions by GEM-H and GEM-P from NWP standpoint. Each seasonal period includes 44 cases where the initialization between two consecutive cases are apart by 36 hours. The first summer and winter cases start at 0000 UTC of 25 June 2014 and 19 December 2014, respectively. The global predictions have been obtained with uniform off-centering in the discretized dynamical equations ($b_m=b_h=b_{nh}=0.2$). \cite{hgi17} have shown that inconsistent off-centering has negligible effects for three-dimensional NWP applications. Therefore, the global forecasts are carried out without any off-centering in the discretized trajectory equations. Furthermore, as has been mentioned in section \ref{s_coupling}, physics is coupled with dynamics through the explicit method for both GEM-H and GEM-P for all the tests presented in this section.

\begin{figure}
\noindent\includegraphics[width=33pc,angle=0]{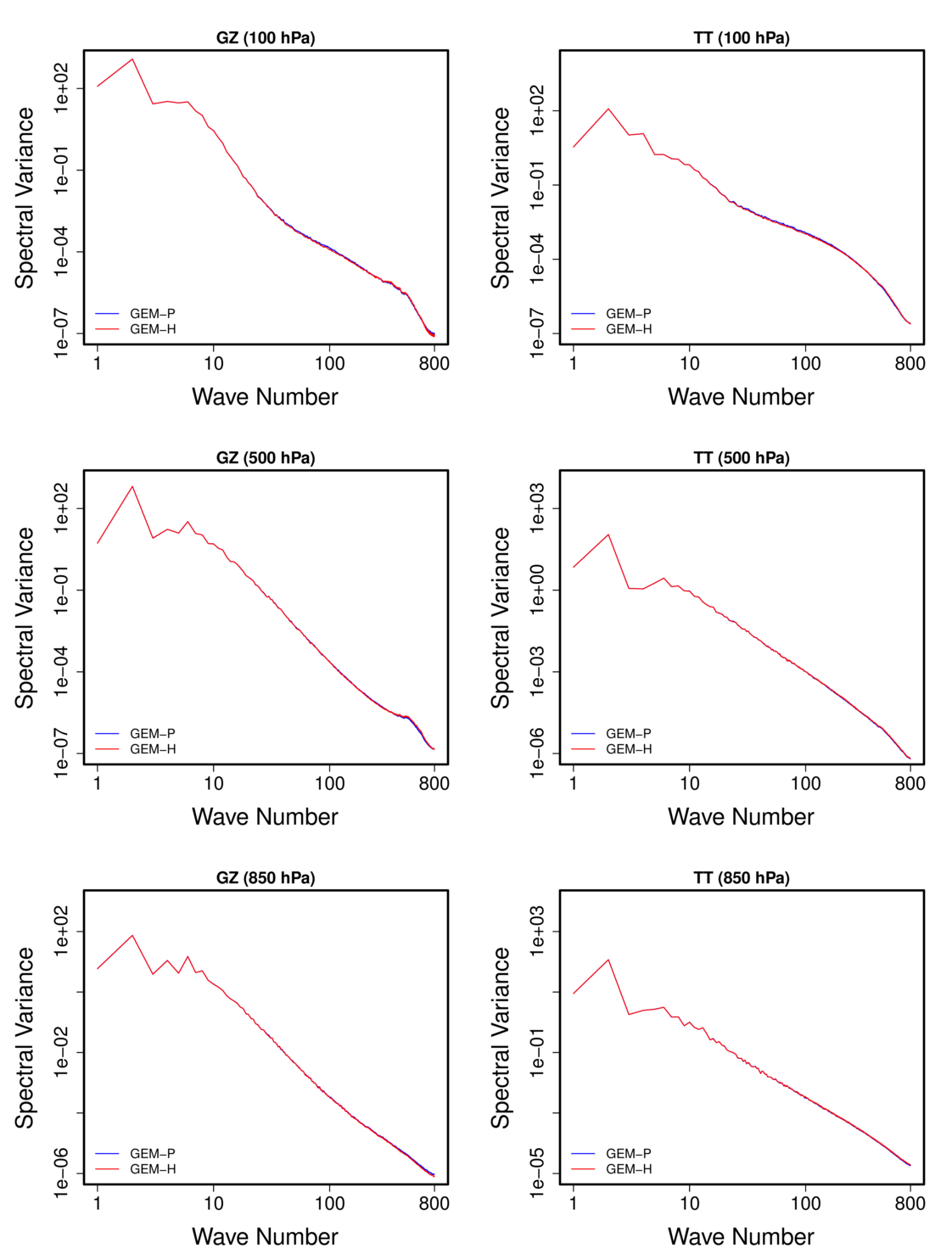}\\
\caption{Spectral variance of (left column) geopotential height (GZ) and (right column) temperature (TT) for 120-h global forecasts with 25 km horizontal grid spacing. Results are presented for three different pressure levels (top: 100 hPa, middle: 500 hPa, bottom: 850 hPa) that are obtained by averaging the spectra for 10 Northern Hemisphere winter forecasts.}
\label{f_spectra_gz}
\end{figure}

\begin{figure}
\noindent\includegraphics[width=33pc,angle=0]{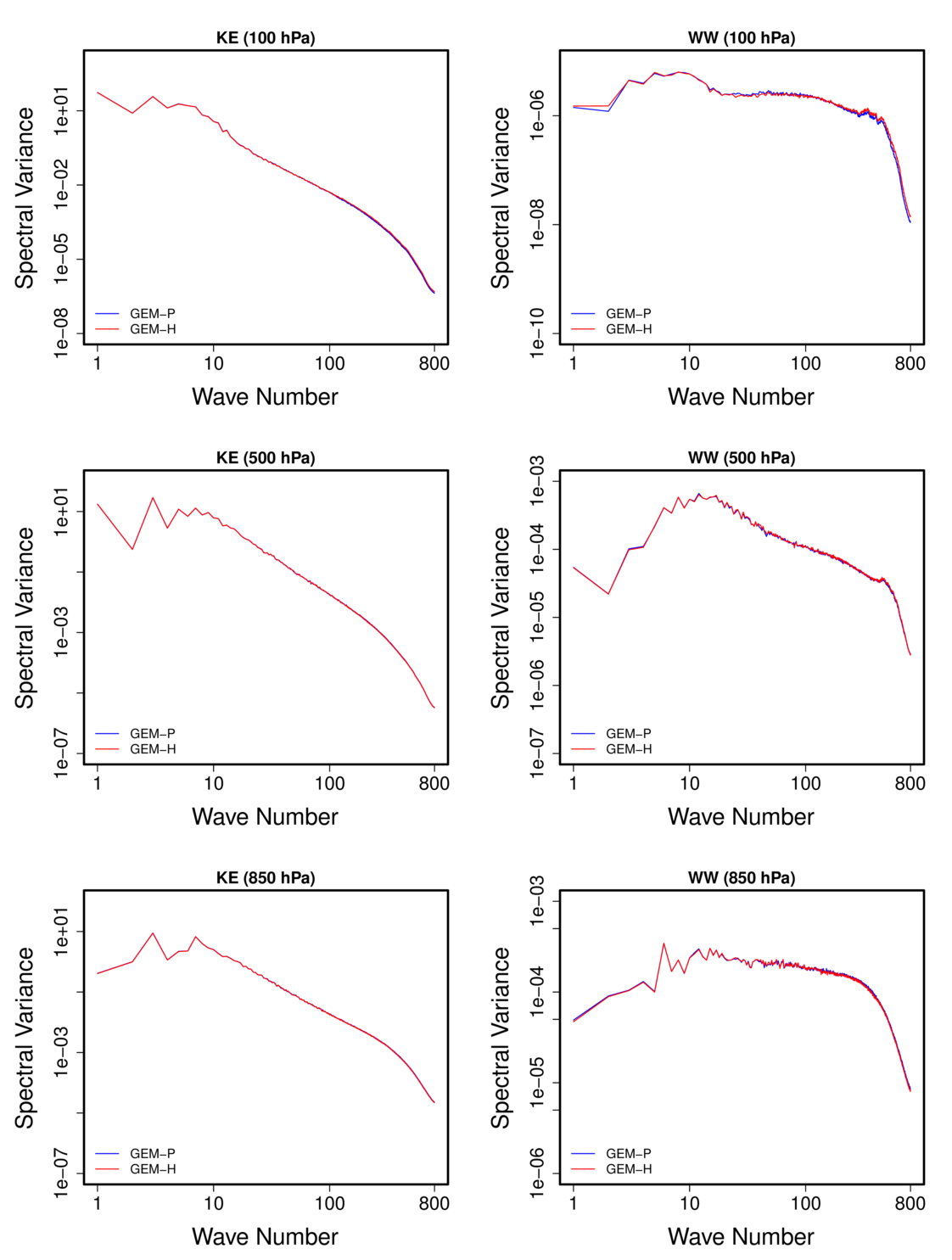}\\
\caption{Spectral variance of (left column) kinetic energy (KE) and (right column) vertical velocity (WW) for 120-h global forecasts. All other conditions as in Fig. \ref{f_spectra_gz}.}
\label{f_spectra_ke}
\end{figure}

First, comparisons are made in the spectral space by comparing the variance spectra associated with the different meteorologically important fields. In order to compute the spectral variance of any meteorological field, it is first interpolated from the Yin-Yang grid to a global Gaussian grid. Afterwards, variance spectra of the field are calculated by decomposing the field at a given pressure level using the spherical harmonics. Figure \ref{f_spectra_gz} shows the spectral variance of the geopotential height and temperature fields for 120-h forecasts at three different pressure levels for an average over 10 cases for the winter period. The results show spectral similarity between GEM-P and GEM-H for the entire range of scales resolved by the global model - from synoptic to mesoscales. The spectra of kinetic energy and vertical velocity are presented in Fig. \ref{f_spectra_ke}. The spectral slope of kinetic energy is critical for accurate representation of atmospheric dynamics, and as   can be seen in this figure, both GEM-H and GEM-P have the same spectral slope at the synoptic and mesoscales. The vertical motion is also important, particularly for physical processes like convection. Fig. \ref{f_spectra_ke} shows close spectral similarity between the vertical motions from the two dynamical cores. This implies that changing the vertical coordinate has negligible sensitivity to the extremely important physical process like convection and convection-driven precipitation. Also, the comparisons in the spectral space confirms that the height-based TFC does not lead to any spurious noise or damping in the meteorological fields for any model-resolved length scale.

\begin{figure}
\noindent\includegraphics[width=45pc,angle=0]{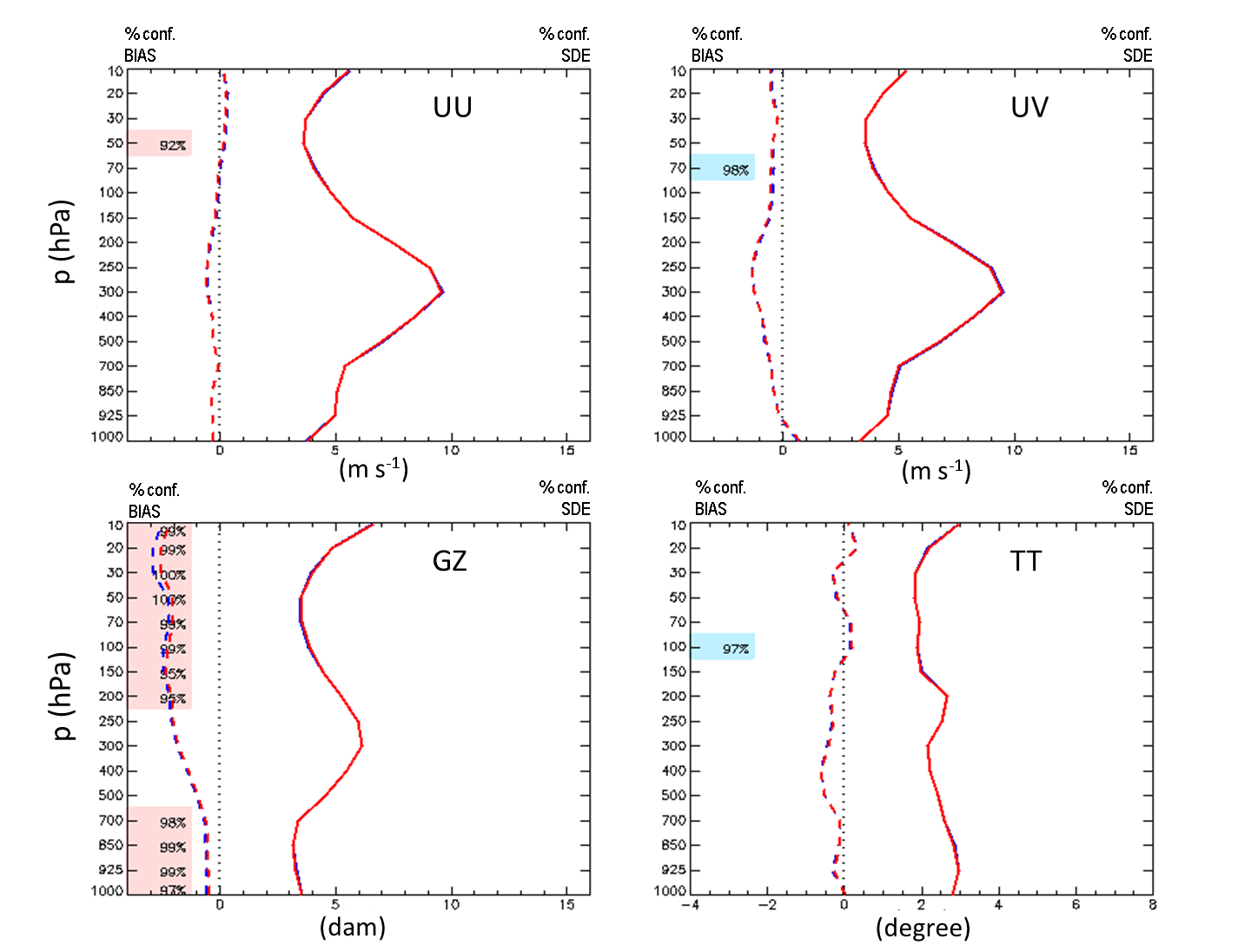}\\
\caption{Comparison of 120-h 25-km GDPS forecasts obtained with GEM-P (blue) and GEM-H (red) dynamical cores against radiosonde observations for zonal wind (UU), wind speed (UV), temperature (TT), and geopotential height (GZ). The dashed and solid lines, respectively, indicate bias and standard deviation of error (SDE). The scores are obtained by averaging over 44 Northern Hemisphere winter cases. The red and blue shaded numbers along the left (right) vertical axes within each panel indicate the confidence in percentage in the statistically significant improvements in bias (SDE) for the dynamical core associated with each color. Significance for bias and SDE are computed using $t$- and $F$-test, respectively.}
\label{f_arcad_1}
\end{figure}

Objective forecast scores are computed by comparing the model predictions against radiosonde observations at different pressure levels. The evaluation is based on the bias and standard deviation of error (SDE) in model predictions for the individual cases as well as for the average of the 44 cases covering each seasonal period. Figure \ref{f_arcad_1} presents the vertical profile of error in the predictions from GEM-P (blue) and GEM-H (red) for the winter period. These figures represent global average scores of 120-hour forecast from 44 winter cases for zonal wind (UU), wind speed (UV), geopotential height (GZ), and temperature (TT). An important thing to note while reading this figure is the presence of the statistical confidence scores at the different pressure levels along the left and right vertical axes of the individual subplots for bias and SDE, respectively. A confidence value (in \%) shaded in blue (red) color implies statistically significant improvement obtained with the GEM-P (GEM-H) core with respect to the other. The confidence score for the average of SDE and bias are estimated by applying the $F$-test and $t$-test, respectively. Figure \ref{f_arcad_1} reveals that although there are small differences in the bias for the average of the 44 winter cases, there is no statistically significant difference in the SDE. When tested in the absence of physics forcings, no statistically significant difference is found between the two dynamical cores in either bias or SDE (not shown). The meteorological fields are interpolated from the TFC of the dynamics to the $\sigma$-coordinate for physics through vertical interpolation which can lead to small differences in the vertical for the different definitions of the TFC. Physical parameterizations can be sensitive to the position of the vertical levels and is apparently responsible for the small bias differences shown in Fig. \ref{f_arcad_1}. Even though small differences in bias are present, the objective scores from GEM-P and GEM-H can be safely assumed to be equivalent as a whole. During the summer period, the two dynamical cores have also been found to be similarly equivalent (not shown). 

\begin{figure}
\noindent\includegraphics[width=45pc,angle=0]{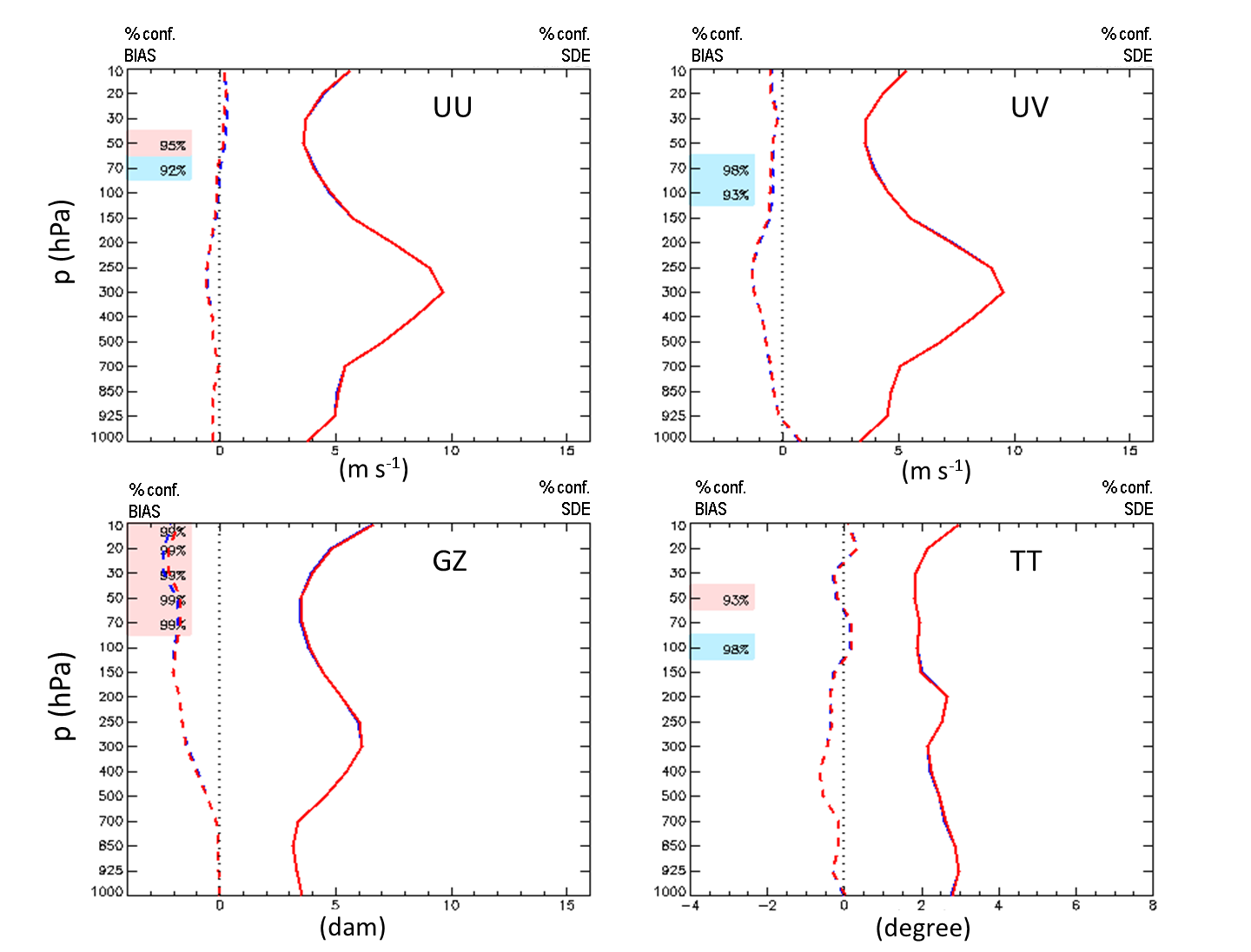}\\
\caption{Same as in Fig. \ref{f_arcad_1}, but with a global mass fixer in the simulations for both GEM-P and GEM-H to improve mass conservation.}
\label{f_arcad_2}
\end{figure}

The geopotential height at 1000 hPa in Fig. \ref{f_arcad_1} shows a negative bias for both dynamical cores. This indicates a loss of mass conservation, which is a consequence of the non-conservative nature of semi-Lagrangian advection. This can be improved by introducing a simple global mass fixer that works by conserving the global mean surface pressure after each dynamics step in the model. The scores in the presence of a global mass fixer is shown in Fig. \ref{f_arcad_2}, where the bias at the lowest model level is eliminated for both dynamical cores. The overall scores for the two cores are again, as expected, found to be equivalent in the presence of a global mass fixer. 

Overall, the results presented in the Figures \ref{f_bubble}-\ref{f_arcad_2} clearly demonstrate that the implementation of GEM-H as presented in this paper produces results that are equivalent to the existing GEM-P dynamical core.

\section{Summary} \label{s_summary}
A newly-developed dynamical core for ECCC's GEM model with a height-based terrain-following vertical coordinate has been presented. With increasing focus on three-dimensional iterative solvers at ECCC driven by the limitations of the operational direct solver as well as the strong numerical instability induced by steep-orography for sub-kilometer resolution NWP, a dynamical core with height-based TCF is expected to be better placed to address the future NWP challenges at ECCC. The principal objective of this paper is to provide information pertaining to the different aspects of the new height-based dynamical core including changes to the model formulation, discretizations, solvers for the discretized problem and the strategy for coupling the new core with the RPN physics package. Another important objective is to demonstrate that the new GEM-H core is capable of making meteorological predictions that are equivalent to the existing GEM-P dynamical core, which is based on a log-hydrostatic-pressure-type vertical coordinate. 

Numerical experiments have been conducted throughout the different stages of GEM-H development. Initially, the bubble convection test revealed that the advection and the buoyancy effects in GEM-H are accurately represented and are producing results that are equivalent to GEM-P. When tested for the idealized Sch\"{a}r's mountain case, the nonhydrostatic and hydrostatic components of the mountain waves predicted by GEM-H are found to be very close to the GEM-P predictions as well as the analytical solution. The dynamics source terms in GEM are averaged over the air parcel trajectories using the trapezoidal method and the calculation of the RHS terms in the dynamical equations are carried out using cubic interpolation. Although it has not been shown explicitly, similarly to GEM-P, in the absence of any off-centering in the discretized dynamical equations, numerical consistency in GEM-H requires a trapezoidal averaging of the source terms in the discretized trajectory equations along  with a cubic interpolation for the wind fields at the departure positions. Furthermore, in the presence of off-centering, the Sch\"{a}r's mountain case shows that the discretized sources terms in the trajectory equations also necessitate off-centered averaging for the sake of consistent numerics in both GEM-H and GEM-P. In general, the knowledge acquired over years regarding the different numerical aspects of the GEM-P dynamical core is proven to be equally applicable to the case of GEM-H. Comparisons between GEM-H and GEM-P for global deterministic predictions are also presented. The results are found to be equivalent in the spectral space confirming that GEM-H does not produce any spurious noise or damping over the model-resolved scales. When compared against upper-air radiosonde observations, except for small differences in bias, GEM-H and GEM-P are found to produce equivalent results.

The rationale behind the choice of the solution approach for the discretized EBV resulting from the GEM-H formulation as well as the method for dynamics-physics coupling is discussed in detail. Although the general structure of the EBV originating form the GEM-H discretized system of equations require an iterative approach, a simplified approach has been devised where the horizontally-variable metric terms---attributable to the vertical-coordinate transformation---are coupled with the nonlinear terms and are treated with nonlinear iterations. This makes the EBV vertically separable and allows the use of a direct solver which is computationally very efficient for the currently operational NWP system configurations. A three-dimensional iterative solver based on FGMRES is also developed for situations when the simplified approach is not feasible. The fact that the GEM-H core can utilize the direct solver approach for global and regional scale model resolutions, eliminates the concern of computational efficiency as far as its implementation in the current and near-future plans for operational NWP systems at ECCC is concerned.

Improving any NWP model is a continuous process. As a stable GEM-H dynamical core is now developed, a number of other relevant issues are currently being studied. The objective is to improve the GEM model in general and the GEM-H dynamical core in particular. One of the most important short-term goal in this regard is to devise a more numerically consistent and accurate coupling between RPN Physics and GEM dynamics which will benefit both the dynamical cores. Also, extensive research is being conducted to develop highly optimized three-dimensional iterative solvers that can be competitive against the operationally-used direct solver while scaling better for very large number of processor cores for the future generations of massively parallel supercomputers. Currently, the Yin-Yang system uses the Schwarz method where the global solution is produced by iteratively solving two elliptic sub-problems for the two sub-domains (Yin and Yang) separately and updating the solutions in the overlapping regions until a certain convergence criteria is satisfied. One promising solution to reduce the execution time of the iterative solver for the Yin-Yang system is to remove the Schwarz iterations and to solve the two elliptic sub-problems as one by using FGMRES \citep{zal12}. Also, pre-conditioners based on other types of methods, e.g., incomplete factorization, block Gauss-Siedel or multigrid method, could be used in order to improve the convergence rate of the FGMRES solver. Finally, on the GEM-H front, another important short-term objective is to improve its numerical stability over steep orography by implementing more accurate numerical approximation of the horizontal gradients in the discretized dynamical equations.\\

%%%%%%%%%%%%%%%%%%%%%%%%%%%%%%%%%%%%%%%%%%%%%%%%%%%%%%%%%%%%%%%%%%%%%
% ACKNOWLEDGMENTS
%%%%%%%%%%%%%%%%%%%%%%%%%%%%%%%%%%%%%%%%%%%%%%%%%%%%%%%%%%%%%%%%%%%%%
%
\acknowledgments
The authors would like to sincerely thank the members of the Numerical Methods Research Group at RPN for all their thoughtful comments and suggestions during the course of GEM-H development. The authors would like to particularly thank Michel Desgagn\'{e}, St\'{e}phane Gaudreault, Rabah Aider and Vivian Lee for their contributions during the implementation of the GEM-H source code. Also, comments from the internal reviewer, Dr. Christopher Subich, have helped to considerably improve the overall presentation of the paper.

%%%%%%%%%%%%%%%%%%%%%%%%%%%%%%%%%%%%%%%%%%%%%%%%%%%%%%%%%%%%%%%%%%%%%
% APPENDIXES
%%%%%%%%%%%%%%%%%%%%%%%%%%%%%%%%%%%%%%%%%%%%%%%%%%%%%%%%%%%%%%%%%%%%%
%

\appendix[A]
\appendixtitle{Nonlinear and RHS components of the discretized equations}

The nonlinear components of (\ref{e_u_1})--(\ref{e_t_1}) and (\ref{e_k_1}), associated with linear components (\ref{e_l_u})--(\ref{e_l_k}), are given by
\begin{equation}
 N_u = -\bigg(f+\frac{tan\phi}{a}u\bigg)\overline{v}^{XY}+ \bigg(\frac{\overline{T_v}^{X\zeta}}{T_*}-1\bigg)\delta_Xq - \bigg(\frac{\overline{T_v}^{X\zeta}}{T_*}-\mathbf{s_i}\bigg)J_X\overline{J^{-1}_\zeta\delta_\zeta q}^{X\zeta},
 \label{e_n_u}
\end{equation}
\begin{equation}
 N_v = \bigg(f+\frac{tan\phi}{a}\overline{u}^{XY}\bigg)\overline{u}^{XY}+ \bigg(\frac{\overline{T_v}^{Y\zeta}}{T_*}-1\bigg)\delta_Yq - \bigg(\frac{\overline{T_v}^{Y\zeta}}{T_*}-\mathbf{s_i}\bigg)J_Y\overline{J^{-1}_\zeta\delta_\zeta q}^{Y\zeta},
 \label{e_n_v}
\end{equation}
\begin{equation}
N_w = \bigg[\bigg(\frac{T_v}{T_*} - \mathbf{s_i}\bigg)J^{-1}_\zeta-\mathbf{s_d}\bigg]\delta_\zeta q - \bigg(\frac{T_v}{T_*}-1\bigg) g\frac{T^{'}_v}{T_v},
 \label{e_n_w}
\end{equation}
\begin{equation}
 N_c=0,
 \label{e_n_c}
\end{equation}
\begin{equation}
 N_T=\frac{1}{\tau_h}\bigg[\ln\bigg(\frac{T_v}{T_*}\bigg)-\frac{T^{'}_v}{T_v}\bigg],
 \label{e_n_t}
\end{equation}
\begin{equation}
 N_z=0
 \label{e_n_k}.
\end{equation}

The corresponding RHS terms, in the absence physics forcing, take the following forms
\begin{equation}
 R_u = \frac{u}{\tau_m}-\beta_m\bigg[-\bigg(f+\frac{tan\phi}{a}u\bigg)\overline{v}^{XY}+ \frac{\overline{T_v}^{X\zeta}}{T_*}\bigg(\delta_Xq - J_X\overline{J^{-1}_\zeta\delta_\zeta q}^{X\zeta}\bigg)\bigg],
 \label{e_r_u}
\end{equation}
\begin{equation}
 R_v = \frac{v}{\tau_m}-\beta_m\bigg[\bigg(f+\frac{tan\phi}{a}\overline{u}^{XY}\bigg)\overline{u}^{XY}+ \frac{\overline{T_v}^{Y\zeta}}{T_*}\bigg(\delta_Yq - J_Y\overline{J^{-1}_\zeta\delta_\zeta q}^{Y\zeta}\bigg)\bigg],
 \label{e_r_v}
\end{equation}
\begin{equation}
R_w = \frac{w}{\tau_{nh}} - \beta_{nh}\bigg[\frac{T_v}{T_*} \bigg(J^{-1}_\zeta\delta_\zeta q - g\frac{T^{'}_v}{T_v}\bigg)\bigg],
 \label{e_r_w}
\end{equation}
\begin{equation}
 R_c=\frac{1}{\tau_h}\bigg(\frac{q}{c_*^2}+\ln{J_\zeta}\bigg)-\beta\bigg[\delta_Xu+\frac{1}{\cos\phi}\delta_Y(\cos\phi v)+\delta_\zeta\dot{\zeta}-\varepsilon\overline{w}^{\zeta}\bigg],
 \label{e_r_c}
\end{equation}
\begin{equation}
 R_T=\frac{1}{\tau_h}\bigg[\ln\bigg(\frac{T_v}{T_*}\bigg)-\frac{\overline{q}^\zeta}{c_{pd}T_*}\bigg] -\beta[\mu w],
 \label{e_r_t}
\end{equation}
\begin{equation}
 R_z=\frac{z-\zeta}{\tau_{nh}}-\beta_{nh}[\dot{\zeta}-w]
 \label{e_r_k}.
\end{equation}

\appendix[B]
\appendixtitle{Deriving the EBV problem}
The LHS of the EBV problem to be solved at every time step in GEM-H is derived by manipulating the discretized system of equations of the form (\ref{e_lnr}). The sequence of operations to derive the EBV are provided below in terms of its linear components:
\begin{equation}
 L^{'}_c=\delta_XL_u+\frac{1}{\cos\phi}\delta_Y(\cos\phi L_v) - \frac{1}{\tau_m}\bigg(L_c-\frac{\ln J_\zeta}{\tau_h}\bigg),
\end{equation}
\begin{equation}
 L^{'}_z=L_z-\frac{z-\zeta}{\tau_{nh}},
\end{equation}
\begin{equation}
 L^{'}_w=L_w+L^{'}_z/\tau_{nh},
\end{equation}
\begin{equation}
 L^{'}_T=L_T+\mu L^{'}_z,
\end{equation}
\begin{equation}
 L^{''}_T=\Gamma (g\tau_h L^{'}_T+L^{'}_w),
\end{equation}
\begin{equation}
 L^{''}_c=L^{'}_c+\frac{\varepsilon}{\tau_m}\overline{L^{'}_z}^\zeta,
\end{equation}
\begin{equation}
 L^{'''}_c=L^{''}_c+\delta_\zeta L^{''}_T-\varepsilon\overline{L^{''}_T}^\zeta.
\end{equation}

Similar manipulations are also applied to the nonlinear and RHS components of the discretized equations to obtain $R^{'}_c$, $N^{'}_c$, $R^{'}_z$, $N^{'}_z$, $R^{'}_w$, $N^{'}_w$, $R^{'}_T$, $N^{'}_T$, $R^{''}_T$, $N^{''}_T$, $R^{''}_c$, $N^{''}_c$, $R^{'''}_c$ and $N^{'''}_c$.

\appendix[C]
\appendixtitle{Back-substitution to obtain the other prognostic variables}
The solution of the EBV computes the unkown $q$. The rest of the variables are then calculated from the discretized dynamical equations through back-substitution in the following sequence:
\begin{equation}
 u=\tau_m\bigg[R_u-N_u-\bigg(\delta_Xq - \mathbf{s_i} J_X\overline{J^{-1}_\zeta\delta_\zeta q}^{X\zeta}\bigg)\bigg]
 \label{e_back_u},
\end{equation}
\begin{equation}
 v=\tau_m\bigg[R_v-N_v-\bigg(\delta_Yq - \mathbf{s_i} J_Y\overline{J^{-1}_\zeta\delta_\zeta q}^{Y\zeta}\bigg)\bigg]
 \label{e_back_v},
\end{equation}
\begin{equation}
\dot{\zeta} = \tau_m \bigg[R^{''}_T - N^{''}_T - \Gamma\bigg((\mathbf{s_i}J^{-1}_\zeta + \mathbf{s_d})\delta_\zeta q - \mu\overline{q}^\zeta\bigg)\bigg],
 \label{e_back_zd}
\end{equation}
\begin{equation}
 w=\dot{\zeta}-L^{'}_z,
 \label{e_back_w}
\end{equation}
\begin{equation}
 T_v=gT_*\bigg[g- \frac{\dot{\zeta}}{\tau_{nh}} + (\mathbf{s_i}J^{-1}_\zeta + \mathbf{s_d})\delta_\zeta q - R^{'}_w -N^{'}_w\bigg]^{-1}.
 \label{e_back_T}
\end{equation}

%%%%%%%%%%%%%%%%%%%%%%%%%%%%%%%%%%%%%%%%%%%%%%%%%%%%%%%%%%%%%%%%%%%%%
% REFERENCES
%%%%%%%%%%%%%%%%%%%%%%%%%%%%%%%%%%%%%%%%%%%%%%%%%%%%%%%%%%%%%%%%%%%%%
\bibliographystyle{ametsoc2014}
\bibliography{references_syed_husain}

%%%%%%%%%%%%%%%%%%%%%%%%%%%%%%%%%%%%%%%%%%%%%%%%%%%%%%%%%%%%%%%%%%%%%
% FIGURES
%%%%%%%%%%%%%%%%%%%%%%%%%%%%%%%%%%%%%%%%%%%%%%%%%%%%%%%%%%%%%%%%%%%%%

\end{document}